\documentclass[aps,prl,twocolumn,superscriptaddress,english]{revtex4}
\usepackage{amssymb}
\usepackage{graphicx}
\usepackage{amsmath}
\usepackage{amsthm}
\usepackage{amsfonts}
\usepackage[T1]{fontenc}
\usepackage[latin9]{inputenc}
\usepackage{array}
\usepackage{multirow}
\usepackage{color}
\usepackage{esint}
\usepackage{bm}
\usepackage{color}
\usepackage{bbm}
\usepackage{hyperref}
\usepackage{babel}
\usepackage{titlesec}

\begin{document}

\global\long\def\id{\mathbbm{1}}
\global\long\def\ui{\mathbbm{i}}
\global\long\def\ud{\mathrm{d}}

\title{Emergent topology and symmetry-breaking order in correlated quench dynamics}

\author{Long Zhang}
\affiliation{International Center for Quantum Materials and School of Physics, Peking University, Beijing 100871, China}
\affiliation{Collaborative Innovation Center of Quantum Matter, Beijing 100871, China.}
\author{Lin Zhang}
\affiliation{International Center for Quantum Materials and School of Physics, Peking University, Beijing 100871, China}
\affiliation{Collaborative Innovation Center of Quantum Matter, Beijing 100871, China.}
\author{Ying Hu}
\affiliation{State Key Laboratory of Quantum Optics and Quantum Optics Devices, Institute of Laser Spectroscopy, Shanxi University, Taiyuan, Shanxi 030006, China}
\affiliation{Collaborative Innovation Center of Extreme Optics, Shanxi University, Taiyuan 030006, China}
\author{Sen Niu}
\affiliation{International Center for Quantum Materials and School of Physics, Peking University, Beijing 100871, China}
\affiliation{Collaborative Innovation Center of Quantum Matter, Beijing 100871, China.}
\author{Xiong-Jun Liu
\footnote{Correspondence author: xiongjunliu@pku.edu.cn}}
\affiliation{International Center for Quantum Materials and School of Physics, Peking University, Beijing 100871, China}
\affiliation{Collaborative Innovation Center of Quantum Matter, Beijing 100871, China.}
\affiliation{Beijing Academy of Quantum Information Science, Beijing 100193, China}
\affiliation{Institute for Quantum Science and Engineering and Department of Physics, Southern University of Science and Technology, Shenzhen 518055, China}


\begin{abstract}
Quenching a quantum system involves three basic ingredients: the initial phase, the post-quench target phase, and
the non-equilibrium dynamics which carries the information of the former two.
Here we propose a dynamical theory to characterize both the topology and symmetry-breaking order in correlated quantum system, through
quenching the Haldane-Hubbard model from an initial magnetic phase to
topologically nontrivial regime. The equation of motion for the complex pseudospin dynamics is obtained
with the flow equation method, with the pseudospin evolution shown to obey a microscopic Landau-Lifshitz-Gilbert-Bloch equation.
We find that the correlated quench dynamics exhibit robust universal behaviors on the so-called band-inversion surfaces (BISs), from which the nontrivial topology
and magnetic order can be extracted. In particular, the topology of the post-quench regime can be characterized
by an emergent dynamical topological pattern of quench dynamics on BISs, which is robust against dephasing and
heating induced by interactions; the pre-quench symmetry-breaking orders is read out from a universal scaling behavior
of the quench dynamics emerging on the BIS, which is valid beyond the mean-field regime.
This work opens a way to characterize both the topology and symmetry-breaking orders by correlated quench dynamics.
\end{abstract}

\maketitle

Quenching a quantum system across a phase transition, the induced far-from-equilibrium dynamics
carries the information of both the initial and final phases.
Quantum quench has been extensively applied to study non-equilibrium physics from the real-time dynamics~\cite{Polkovnikov2011,Eisert2015,Giamarchi_book,Heyl2018}.
In condensed matter physics, the melting or creation of long-range order can be investigated in the dynamics of symmetry-breaking states,
e.g. the survival of magnetic order following an interaction quench in the Hubbard models~\cite{Tsuji2013,Sandri2013,Balzer2015}.
For topological systems, characterization of topology by quench dynamics has also attracted particular interest very recently~\cite{Tarnowski2017,Song2017,Sun2018,Wang2017,Zhanglin2018,Zhanglong2018,Yu2018,McGinley2018a,McGinley2018b}.

So far the dynamical characterization theories are applicable to noninteracting topological systems~\cite{Tarnowski2017,Sun2018,Song2017,Wang2017,Zhanglin2018,Zhanglong2018,Yu2018,McGinley2018a,McGinley2018b}.
For an interacting system, the more challenging but interesting issues could arise.
First, the single-particle quantum numbers in correlated systems are not conserved.
It is unclear how to define the dynamical topology for the characterization.
Second, the interaction can bring about complex effects~\cite{Polkovnikov2011}, such
as dephasing and heating. Their influence on topology remains an open question.
Third, symmetry-breaking orders can emerge in correlated systems.
The compositive characterization of both
topology and symmetry-breaking orders in a correlated
quantum quench is, at present, an outstanding issue.

\begin{figure}
\includegraphics[width=0.49\textwidth]{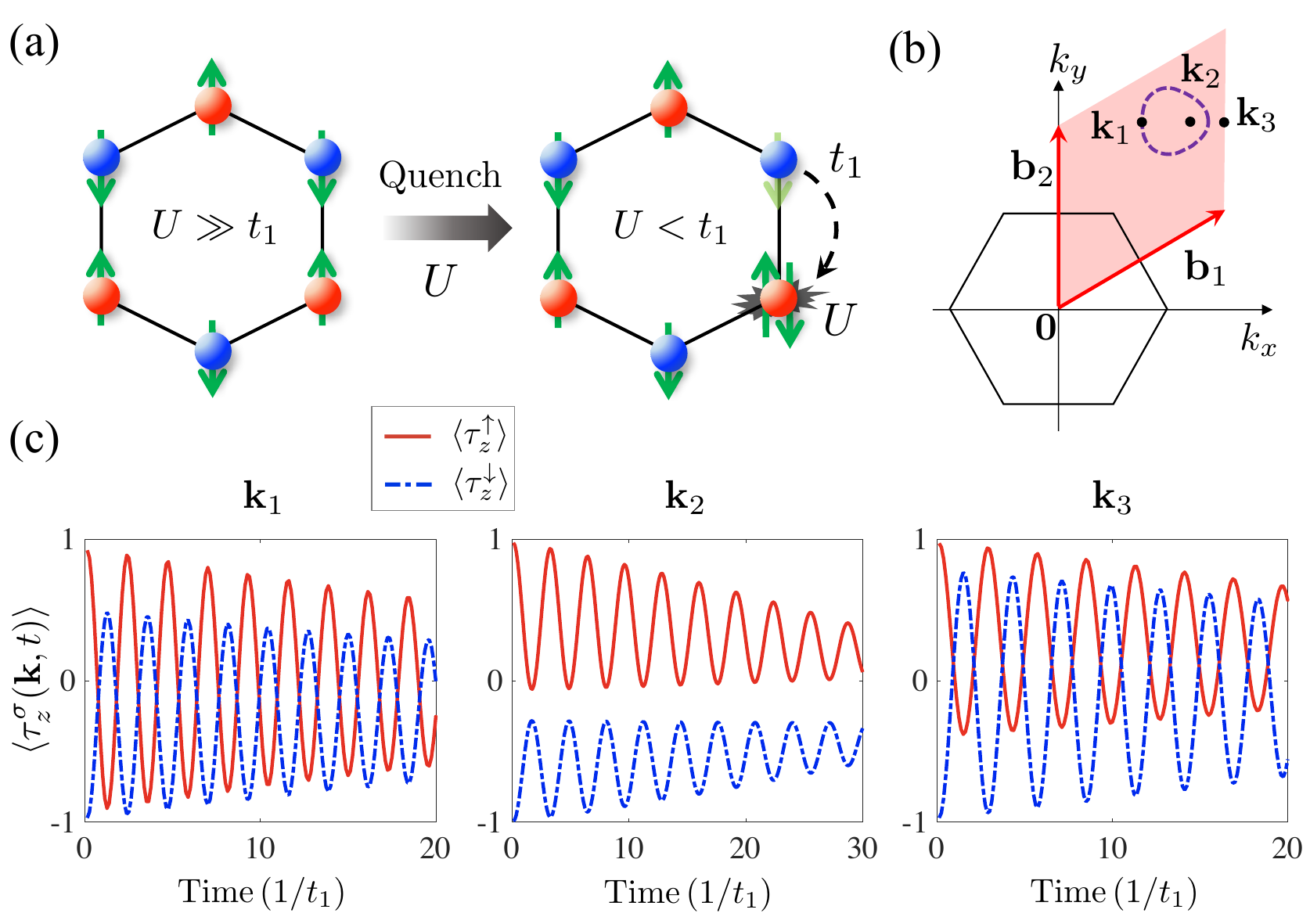}
\caption{Interaction quench and pseudospin dynamics. (a) The system undergoes a transition from an AF phase to a topologically
nontrivial phase by quenching the interaction from $U\gg t_1$ to $U<t_1$. (b) The first Brillouin zone (hexagon) with
the reciprocal-lattice vectors ${\bf b}_{1}=\frac{2\pi}{3a_0}(\sqrt{3},1)$ and ${\bf b}_2=\frac{4\pi}{3a_0}(0,1)$ ($a_0$ is the lattice constant).
The dashed purple line denotes the band-inversion surface of spin-up component. (c) The pseudospin polarization $\langle\tau_z^\sigma\rangle$
oscillates after the quench for each spin $\sigma=\uparrow\downarrow$. Three points in the Brillouin zone (b) are taken for example.
Here $t_2=0.3t_1$, $M=-0.5t_1$, $m_{\rm C}=0.5t_1$, $m_{\rm AF}=4t_1$, and $U=0.3t_1$ after quench.
}\label{fig1}
\end{figure}

We propose for the first time a dynamical theory to characterize both topology and symmetry-breaking orders, through quenching the spin-$1/2$ Haldane model~\cite{Haldane1988,Jotzu2014} with Hubbard interaction from an initial magnetic phase (Fig.~\ref{fig1}a), which exists in strongly interacting regime~\cite{Hickey2015,Zheng2015,Liu2016,Arun2016,Hickey2016,Vanhala2016,JXWu2016,Rubio2018}, to topological regime with relatively weak interaction.
We show that the particle-particle interaction has nontrivial correlation effects on the pseudospin dynamics which,
after being projected onto the momentum space, follow a novel microscopic Landau-Lifshitz-Gilbert-Bloch equation.
With this the dephasing and heating effects are explicitly predicted. Importantly, we find that the correlated quench dynamics
exhibit emergent robust topological pattern and universal scaling on the one-dimensional (1D) momentum subspaces
called band-inversion surfaces (BISs)~\cite{Zhanglin2018,Zhanglong2018}.
These exotic features manifest a deep dynamical bulk-surface correspondence relating both topology and symmetry-breaking orders to correlated quench dynamics on BISs.

{\em The model.---}The full Hamiltonian of the 2D Haldane-Hubbard model with onsite interaction $U$ reads
\begin{align}\label{H1}
H&=H_0+U\sum_{\vec i}(a^\dagger_{\vec i\uparrow}a^\dagger_{\vec i\downarrow}a_{\vec i\downarrow}a_{\vec i\uparrow} +b^\dagger_{\vec i\uparrow}b^\dagger_{\vec i\downarrow}b_{\vec i\downarrow}b_{\vec i\uparrow}),\\
H_0&=-t_1\sum_{\langle \vec i\vec j\rangle,\sigma}(a^\dagger_{\vec i\sigma}b_{\vec j\sigma}+{\rm h.c.})-t_2\sum_{\langle\langle \vec i\vec j\rangle\rangle,\sigma}(e^{\ui\phi}a^\dagger_{\vec i\sigma}a_{\vec j\sigma}
\nonumber\\
&+e^{-\ui\phi}b^\dagger_{\vec i\sigma}b_{\vec j\sigma}+{\rm h.c.})+M\sum_{\vec i,\sigma}(a^\dagger_{\vec i\sigma}a_{\vec i\sigma}-b^\dagger_{\vec i\sigma}b_{\vec i\sigma}).\nonumber
\end{align}
Here $a_{\vec i\sigma }$ $(b_{\vec i\sigma})$ and $a_{\vec i\sigma}^\dag$ ($b_{\vec i\sigma}^\dag$) are annihilation and creation operators, respectively, for fermions of
spin $\sigma=\uparrow,\downarrow$ on $A$ ($B$) sites. The nearest- ($t_1$) and next-nearest-neighbor ($t_2$) hopping is considered,
with the latter having a phase $\pm\phi$. $M$ is an energy imbalance between $A$ and $B$ sites.

The noninteracting Hamiltonian 
$H_0=\sum_{\bf k,\sigma}{\bf h}({\bf k})\cdot{\bm\tau}^\sigma$,
where ${\bf h}({\bf k})=(h_x,h_y,h_z)$ mimics an effective Zeeman field in Bloch $\bold k$ space~\cite{Suppl}, with
the pseudospin operators $\tau_{z}^\sigma=a_{{\bf k}\sigma}^\dagger a_{{\bf k}\sigma}-b_{{\bf k}\sigma}^\dagger b_{{\bf k}\sigma}$,
$\tau_{x}^\sigma=a_{{\bf k}\sigma}^\dagger b_{{\bf k}\sigma}+{\rm h.c.}$, and
$\tau_{y}^\sigma=-\ui[\tau_{z}^\sigma, \tau_{x}^\sigma]$.
It has been widely studied~\cite{Hickey2015,Zheng2015,Liu2016,Arun2016,Hickey2016,Vanhala2016,JXWu2016,Rubio2018} that in the ground state $|\Psi_G\rangle$, an antiferromagnetic (AF) order $m_{\rm AF}=(m_\downarrow-m_\uparrow)/2$
arises for strong repulsive interaction, and the energy imbalance $M$ further leads to
a charge order $m_{\rm C}=(m_\uparrow+m_\downarrow)/2$ corrected by Hubbard interaction, characterizing the population difference in the two sublattices. Here the initial orders $m_\sigma\equiv \langle a^\dagger_{\vec i\sigma}a_{\vec i\sigma}-
b^\dagger_{\vec i\sigma}b_{\vec i\sigma}\rangle U_{\rm in}/4$, with $U_{\rm in}$ the initial strong interaction, and the expectation $\langle\cdot\rangle$ computed in ground state $|\Psi_G\rangle$. For the quench study, we can write down initial ground state in the mean-field form $|\Psi_G\rangle\rightarrow|\Psi_{\rm MF}\rangle$ which solely depends on the order parameters~\cite{Suppl}, or as the Gutzwiller many-body wave function which is beyond mean field picture~\cite{Gutzwiller_S,Li1993_S,Eichenberger2007_S}, and investigate its evolution under the post-quench Hamiltonian with relatively weak interactions. As studied below and detailed in Supplementary Material~\cite{Suppl}, the central results in this work are valid beyond mean-field regime.

We solve the quench dynamics by the flow equation method~\cite{Wilson1993,Wilson1994,Wegner1994}.
The process is below. First, through a unitary transformation that changes continuously with a flow parameter $l$,
we (nearly) diagonalize the post-quench Hamiltonian at $l\to\infty$~\cite{Kehrein_book}. Accordingly, the transformation of an
operator ${\cal O}(l)$ (including the Hamiltonian) follows the flow equation $d{\cal O}(l)/dl=[\eta(l),{\cal O}(l)]$,
where the canonical generator $\eta(l)=[H_0(l),H_I(l)]=-\eta(l)^\dagger$ is anti-Hermitian, with
$H_I$ the interacting term of the full Hamiltonian. Second, the time-evolved operator ${\cal O}(l\to\infty,t)$ is
obtained straightforwardly in the diagonal bases. Finally, we perform the backward transformation so that the operator
flows back as ${\cal O}(l\to\infty,t)\rightarrow{\cal O}(0,t)$~\cite{Moeckel2008,Moeckel2009}.
The time evolution is then given in the original bases.

We apply this method to the present system (details are given in supplementary material~\cite{Suppl}).
We consider the ansatz below for post-quench regime
\begin{align}\label{Hl}
H(l)&=\sum_{{\bf k},\sigma,s=\pm}{\cal E}_s({\bf k}):c_{{\bf k},s\sigma}^\dagger c_{{\bf k},s\sigma}:+\nonumber\\
&\sum_{\substack{{\bf p'pq'q}\\ s_1s_2s_3s_4}}U_{\bf p'pq'q}^{s_1s_2s_3s_4}(l):c_{{\bf p'},s_1\uparrow}^\dagger c_{{\bf p},s_2\uparrow}
c_{{\bf q'},s_3\downarrow}^\dagger c_{{\bf q},s_4\downarrow}:,
\end{align}
where ${\cal E}_\pm({\bf k})$ are the band energies of $H_0$, the normal ordering is with respect to the initial state $|\Psi_G\rangle$,
and $c_{{\bf k},\pm\sigma}^\dagger$ ($c_{{\bf k},\pm\sigma}$) are the creation (annihilation) operators of
spin $\sigma=\uparrow\downarrow$ for the upper and lower band states of $H_0$~\cite{Suppl}.
The interaction strength $U_{\bf p'pq'q}^{s_1s_2s_3s_4}(l)$ is defined for momentum-conserved scattering channels. With the interaction weaker than the band width, only the leading order contributions from scatterings up to $U^2$ will be considered. Taking the previously defined canonical generator $\eta(l)$, the interaction $U(l)$ decays exponentially with $l$ and flows to zero at $l\rightarrow\infty$.
We then work out the flow of creation and annihilation operators with respect to $A$ and $B$ sites, with
${\cal A}_{{\bf k}\uparrow}^\dagger(l=0)=a_{{\bf k}\uparrow}^\dagger$ and ${\cal B}_{{\bf k}\uparrow}^\dagger(l=0)=b_{{\bf k}\uparrow}^\dagger$,
from the same generator $\eta(l)$. Finally we obtain time evolution of pseudospin polarization at momentum $\bold k$, calculated by
$\langle\tau^{\sigma}_z({\bf k},t)\rangle=\langle\Psi_G|{\cal A}^\dagger_{{\bf k}\sigma}(l=0,t){\cal A}_{{\bf k}\sigma}(l=0,t)
-{\cal B}^\dagger_{{\bf k}\sigma}(l=0,t){\cal B}_{{\bf k}\sigma}(l=0,t)|\Psi_G\rangle$, similar for $\langle\tau^{\sigma}_{x,y}({\bf k},t)\rangle$.
Note that the single-particle $\bold k$ is no longer conserved. We project the results onto the single-particle
momentum space to study the pseudospin dynamics (Fig.~\ref{fig1}b,c).

\begin{figure*}
\includegraphics[width=0.7\textwidth]{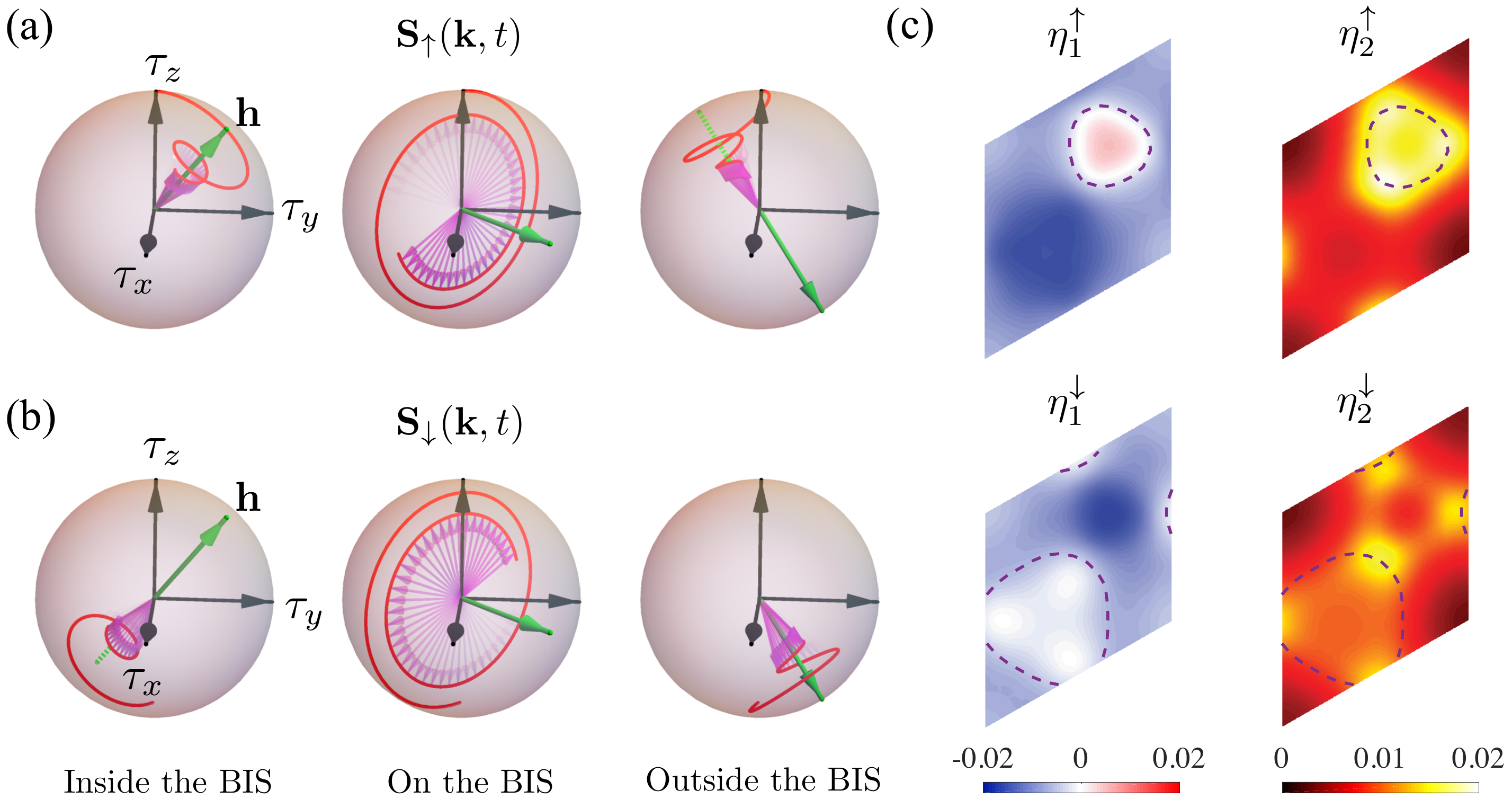}
\caption{Pseudospin dynamics from equation of motion.
(a-b) Time evolution of pseudospin vectors for spin-up (a) and spin-down (b). Damping and heating
effects exhibit features in different regions with respect to the BIS.
(c) The calculated distribution of damping factors $\eta_{1}^\sigma$ and heating
factors $\eta_{2}^\sigma$. The dashed purple lines denotes the {\em noninteracting} BISs for each spin.
Here we take $t_2=0.3t_1$, $M=-0.5t_1$, $m_{\rm C}=0.5t_1$, $m_{\rm AF}=4t_1$, and the interaction $U=0.3t_1$ after quench.
}\label{fig2}
\end{figure*}

{\em Equation of motion for pseudospin dynamics.---}
We show that the essential physics of pseudospin dynamics can be captured by the equation of motion
in the projected $\bold k$-space with ${\bf S}_{\sigma}({\bf k},t)\equiv\frac{1}{2}\langle{\bm\tau}^\sigma({\bf k},t)\rangle$. Taking into account the leading-order contributions we obtain~\cite{Suppl}
\begin{equation}\label{EOM}
\frac{d{\bf S}_{\sigma}(t)}{dt}={\bf S}_{\sigma}(t)\times2{\bf h}-\eta_1^\sigma{\bf S}_{\sigma}(t)\times\frac{d{\bf S}_{\sigma}(t)}{dt}
-\eta_2^\sigma\frac{{\bf S}_{\sigma}(t)}{T_{\rm g}},
\end{equation}
where the first $2\bold h$-term corresponds to the single-particle precession, and the second and third terms are induced by correlation effects. The $\eta_{1}^\sigma$-term
represents the interaction induced damping of precession, and $\eta_{2}^\sigma$-term leads to dephasing and heating,
with $T_{\rm g}\equiv1/(2E_0)$ and $E_0({\bf k})=[h^2_x({\bf k})+h^2_y({\bf k})+h^2_z({\bf k})]^{1/2}$.
This equation renders a novel mixed {\it microscopic} form of Landau-Lifshitz-Gilbert~\cite{Gilbert2004} and Bloch
equations~\cite{Bloch1946} for magnetization, and is not altered in characterizing the initial phase in mean-field theory or as Gutzwiller state, albeit the beyond-mean-field effects can correct the coefficients of the equation~\cite{Suppl}.
The solution reads generically
\begin{align}\label{St}
{\bf S}_{\sigma}(t)={\bf S}_{\sigma}^{(0)}+{\bf S}_{\sigma}^{(c)}(t)+{\bf S}_{\sigma}^{(h)}(t)+{\bf S}_{\sigma}^{(l)}(t),
\end{align}
where ${\bf S}_{\sigma}^{(0)}({\bf k})=\delta n_\sigma({\bf k}){\bf h}({\bf k})/E_0({\bf k})$ is the incoherent time-independent part,
with $\delta n_\sigma({\bf k})=n^\sigma_{+-}({\bf k})-n^\sigma_{--}({\bf k})$ being the density difference of the initial state populated
in the upper ($n^\sigma_{+-}$) and lower ($n^\sigma_{--}$) eigen-bands,
${\bf S}_{\sigma}^{(c)}({\bf k},t)\sim\cos(t/T_{\rm g})$
is the single-particle coherent oscillation, ${\bf S}_{\sigma}^{(h)}({\bf k},t)\approx-\lambda^{\sigma}_1({\bf k},t){\bf S}_{\sigma}^{(c)}({\bf k},t)$
($\lambda^{\sigma}_1\propto U^2/E_0^2$) represents the interaction-induced high-frequency fluctuation, which reduces the
single-particle procession, and ${\bf S}_{\sigma}^{(l)}({\bf k},t)\approx-2\lambda^{\sigma}_2({\bf k},t){\bf h}({\bf k})/E_0({\bf k})$
($\lambda^{\sigma}_2\propto U^2/E_0^2$) denotes the low-frequency interaction effect, which equilibrates the density distribution
on upper and lower bands (heating). The coefficients $\lambda_{1,2}^\sigma$ are related to the factors $\eta_{1,2}^\sigma$ (see later).
Note that the entire many-body system evolves unitary. The dephasing and heating arise in the projected quench dynamics
at fixed $\bold k$, since all the particles with $\bold k'\neq\bold k$ act as a bath scattering the $\bold k$ state.

The $\eta_{1,2}^\sigma$ terms are momentum dependent. For comparison, we first define
the BIS for single-particle Hamiltonian $H_0$~\cite{Zhanglin2018,Zhanglong2018,Sun2018}, being the momentum subspace where band inversion occurs and time-averaged
spin polarizations $\overline{\bold S_\sigma(\bold k,t)}|_{U=0}=0$, equivalent to ${\bf S}_{\sigma}^{(0)}({\bf k})=0$.
On the single-particle BIS, we have $\eta_1^\sigma\simeq-4(d\lambda^{\sigma}_2/dt)T_{\rm g}$ and
$\eta_2^\sigma\simeq4(d\lambda^{\sigma}_1/dt)T_{\rm g}n^\sigma_{+-}n^\sigma_{--}$, where
$d\lambda^{\sigma}_{1,2}/dt$ are approximately constant in early time~\cite{Suppl}.
Fig.~\ref{fig2}c shows that near the BIS (dashed line) $\eta_1^\sigma$ is small
(due to the cancelling of the two-band contributions) and the heating due to $\eta_2^\sigma$-term
dominates the correlation effect. In comparison, the damping enhances at $\bold k$ away from BIS.
The heating shortens the pseudospin vector while the damping drags the vector towards the $\bold h$ axis (see Fig.~\ref{fig2}a-b).

{\em Topology emerging on BIS.--}We show now the correlated pseudospin dynamics on BISs exhibiting emergent topological pattern, which corresponds to the post-quench topology. From Eq.~\eqref{EOM} one finds that the damping
$\eta_1$-term modifies the procession. Thus the BISs in the presence of interactions, with $\overline{\bold S_\sigma(\bold k,t)}|_{U\neq0}=0$,
is deformed from the single-particle BISs where $\bold h$ is perpendicular to ${\bold S}_\sigma$. In contrast, one can prove that the positions $\bold k=\bold k_\rho$
of topological charges, with $\bold h_{\rm so}(\bold k_\rho)\equiv(h_{y},h_{x})=0$ in the noninteracting regime, is unchanged from the equation of motion~\eqref{EOM}.
As shown below, the emergent topology of quench dynamics on BISs reflects the total topological charges enclosed by the BISs.

\begin{figure}[h]
\includegraphics[width=0.48\textwidth]{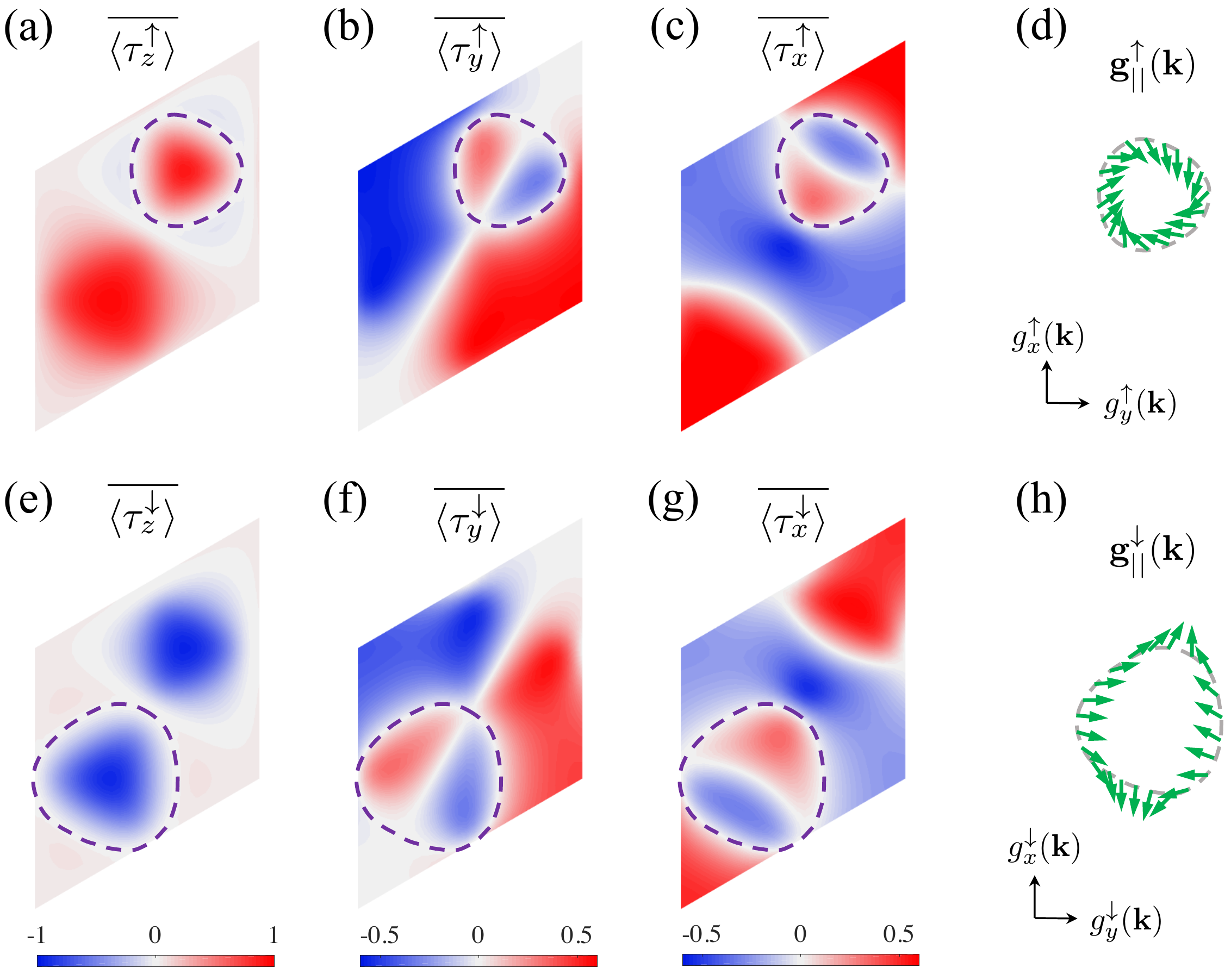}
\caption{
Emergent topology of quench dynamics. Time-averaged pseudospin polarizations $\overline{\langle\tau_{x,y,z}^\sigma({\bf k},t)\rangle}$ for
spin-up (a-b) and spin-down (e-g) with the corresponding dynamical fields ${\bf g}_{\parallel}^\sigma({\bf k})$ (d,h). The dashed lines denotes the BISs.
The constructed dynamical field on the BIS for either spin characterizes the topology with Chern number $C=1$. Here we take $t_2=0.3t_1$,
$M=-0.5t_1$, $m_{\rm C}=0.5t_1$ and $m_{\rm AF}=4t_1$,
and the post-quench interaction $U=0.3t_1$.  The time average is taken over 5 times of oscillation period for each ${\bf k}$.
}\label{fig3}
\end{figure}

To characterize the topology emerging on the BISs, we introduce a dynamical field ${\bf g}^\sigma({\bf k})$, with the components $\bold g^\sigma(\bold k)=\pm\frac{1}{{\cal N}_k}\partial_{k_\perp}\overline{\bold S_\sigma(\bold k,t)}$. It takes $+$ (or $-$) for
$\sigma=\uparrow$ (or $\downarrow$), the momentum $k_\perp$ is
perpendicular to the BIS, and ${\cal N}_k$ is the normalization factor. Due to the damping and heating effects, the ${\bf g}^\sigma({\bf k})$ vector is generally
not in the $x$-$y$ plane. To characterize the topology, we project the dynamical field onto the $x$-$y$ plane such that
${\bf g}_{\parallel}^\sigma({\bf k})=\hat e_\parallel\cdot{\bf g}^\sigma({\bf k})=(g^\sigma_y,g^\sigma_x)$,
and can prove that ${\bf g}^\sigma_\parallel({\bf k})\simeq{\bf h}_{\rm so}({\bf k})$ on the interacting BISs~\cite{Suppl}.
Therefore, the winding of ${\bf g}^\sigma_\parallel({\bf k})$ on BISs quantifies the total topological charges (at zeros of $\bold h_{\rm so}$-vector) enclosed by the BISs, corresponding to the topology of the post-quench regime and
valid for the present interacting system. This new characterization is different from the free-fermion regime, where the topology emerges
in the bare dynamical field $\bold g^\sigma(\bold k)$~\cite{Zhanglin2018,Zhanglong2018}, not directly applicable to the regime with interactions.
A typical example for the topology of the dynamical field is illustrated in Fig.~\ref{fig3}d,h.

{\em Magnetic order from quench dynamics on BIS.--}The AF and charge orders are closely related to the spin and density distributions in $A$ and $B$ sites, hence related to the pseudospin dynamics, in which the
BISs also play the pivotal role. For the initial phase characterized by the mean-field theory that $|\Psi_G\rangle\rightarrow|\Psi_{\rm MF}\rangle$,
the BIS defined by $\overline{{\bf S}_{\sigma}(\bold k,t)}=0$ is alternatively interpreted as the momenta
satisfying
$E_0^2({\bf k})+m_{\sigma}h_z({\bf k})=-(d\lambda^\sigma_{2}/dt)TE_0({\bf k})E_0^\sigma({\bf k})$ with $E_0^\sigma\equiv\sqrt{E_0^2+2m_\sigma h_z+m_\sigma^2}$.
Here $T$ is the interval for time averaging and the right-hand side represents shift of BISs by interaction.
This formula shows that BISs depends on both the pre-quench phase ($m_{\sigma}$)
and the post-quench Hamiltonian. Further, the half amplitude,
defined as $Z_{0}^\sigma({\bf k})\equiv\langle\tau_z^\sigma({\bf k},t=0)\rangle$,
on BISs reads
$Z^\sigma_0=(d\lambda^\sigma_{2}/dt)Th_z/E_0-m_\sigma(E_0^2-h_z^2)/(E_0^2E_0^\sigma)$,
which also relates the magnetization to quench dynamics.
With these results and up to the leading order correction from interaction, we show the scaling~\cite{Suppl}
\begin{equation}\label{mag}
f(m_\sigma)=-\frac{{\rm sgn}(Z^\sigma_0)}{g(Z_0^\sigma)}+O(U^4),
\end{equation}
where $f(m_\sigma)=m_\sigma T_0$ and $g(Z^\sigma)=\sqrt{1-Z^{\sigma\,2}_0}/\pi$, with $T_0({\bf k})=\pi/E_0$.
The result in Eq.~(\ref{mag}) gives a universal scaling at any $\bold k$ on BISs, insensitive to interactions.

\begin{figure}
\includegraphics[width=0.45\textwidth]{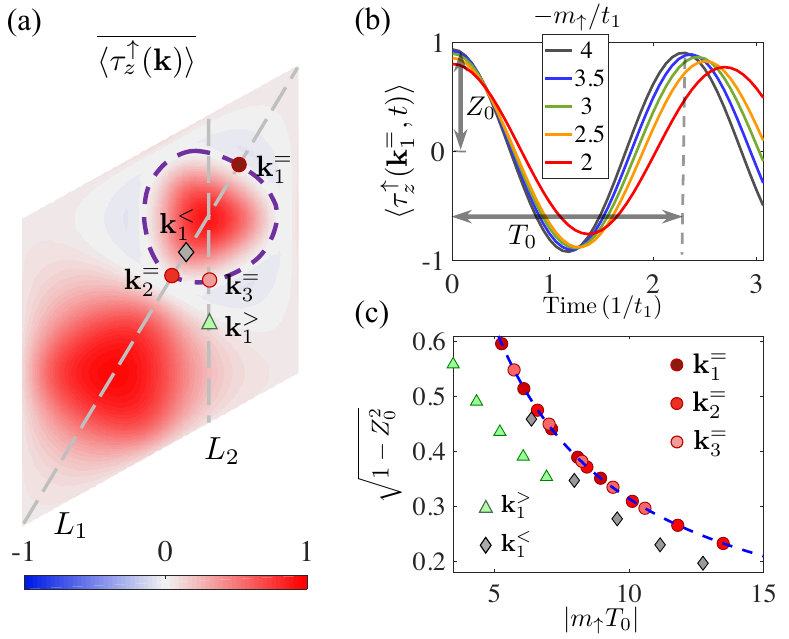}
\caption{Characterizing symmetry-breaking order. (a) The momenta taken for measurement.
Three points lie on the BIS, with ${\bf k}^=_{1,2}$ being in the line $L_1$: $(k_x,k_y)={\bf b}_1+{\bf b}_2$ and
${\bf k}^=_{3}$ in the line $L_2$: $k_x=\frac{4\pi}{9a_0}\sqrt{3}$.
One is chosen inside the BIS with ${\bf k}^<_1=\frac{3}{5}({\bf b}_1+{\bf b}_2)$, and one is outside the BIS with
${\bf k}^>_1=\frac{2}{3}{\bf b}_1+\frac{1}{3}{\bf b}_2$.
(b) Both the half oscillation amplitude $Z_0$ and the oscillation period $T_0$ are
measured for different magnetization $m_{\uparrow}/t_1=-\{2,2.5,3,3.5,4\}$.
(c) The results are shown as $(|m_{\uparrow}T_0|,\sqrt{1-Z_0^2})$. The data taken
on the BIS all satisfy the function $f(x)=\pi/x$ (dashed curve), which verifies the relation Eq.~(\ref{mag}).
Here we take $t_2=0.3t_1$, $M=-0.5t_1$ and the post-quench interaction $U=0.3t_1$.
}\label{fig4}
\end{figure}

We provide numerical results in Fig.~\ref{fig4}a for spin-up component. By identifying the BIS (the dashed purple curve), we record the short-time dynamics at momenta of three kinds: inside (${\bf k}^<_1$), outside (${\bf k}^>_1$), and right on the BIS (${\bf k}^=_{1,2,3}$).
We measure both $Z_0(\bold k)$ and $T_0(\bold k)$ versus order parameters $m_{\uparrow}/t_1$ (Fig.~\ref{fig4}b).
The results are plotted as points $(|m_\uparrow T_0|,\sqrt{1-Z_0^2})$ in Fig.~\ref{fig4}c, showing that
the data measured on the BIS all satisfy the scaling~(\ref{mag}). In experiment, one can obtain $m_\sigma$ by measuring only the first one or two oscillations. The AF order is then obtained by
$m_{\rm AF}=(m_\downarrow-m_\uparrow)/2$, and the charge order is $m_{\rm C}=(m_\uparrow+m_\downarrow)/2$. Finally, we emphasize that the scaling~(\ref{mag}) is satisfied beyond the mean-field theory. For the initial phase described with the correlated Gutzwiller wave function $|\Psi_G\rangle$, the same scaling holds, with $Z_0^{\sigma}$ and the order parameters $m_\sigma$ renormalized by correlations in the more precise Gutzwiller ground state. For simplicity we put the details in the Supplementary Material~\cite{Suppl}.

{\em Conclusion.--}We have proposed a dynamical theory to characterize both the topology and symmetry-breaking orders by quantum quench in correlated topological system. By quenching the Haldane-Hubbard model from initial symmetry-breaking phase into topological regime, the induced quantum dynamics on band inversion surfaces (BISs) exhibits emergent topology and universal scaling, which uniquely correspond to and thus characterize the post-quench equilibrium topological state and pre-quench symmetry-breaking orders, respectively. Our results are shown to be valid beyond mean-field theory~\cite{Suppl}, hence reveal a deep dynamical bulk-surface correspondence for topology and symmetry-breaking orders.
These results are expected to be generically seen in Chern-Hubbard insulators and 1D topological-Hubbard systems. Note that the pseudospin
dynamics can be measured by the tomography of Bloch states~\cite{Hauke2014,Flaschner2016}.
This work opens an avenue to explore profound correlation physics with novel topology by quench dynamics.


This work was supported by the National Key R\&D Program of China (2016YFA0301604, 2017YFA0304203), National Natural Science Foundation
of China (11574008, 11761161003, 11825401, and 11874038), and the Strategic Priority Research Program of Chinese Academy of Science
(Grant No. XDB28000000). Y.H. also acknowledges support from the National Thousand-Young-Talents Program, and Changjiang Scholars
and Innovative Research Team (Grant No. IRT13076).


\setcounter{equation}{0} \setcounter{figure}{0}
\setcounter{table}{0} 
\renewcommand{\theparagraph}{\bf}
\renewcommand{\thefigure}{S\arabic{figure}}
\renewcommand{\theequation}{S\arabic{equation}}

\onecolumngrid
\flushbottom
\newpage

\section*{\normalsize SUPPLEMENTAL MATERIAL}

In the Supplementary Materials, we first provide the details in deriving the flow equation, dynamical topology, and universal scaling on BISs when the initial ground state is characterized with mean-field theory. Then, we show in detail that these results can be further obtained beyond the mean-field theory.

\section{I. Hamiltonians and the mean-field ground state}

The Bloch Hamiltonian of the noninteracting Haldane model regardless of the spin can be written as
\begin{equation}
{\cal H}({\bf k})={\bf h}({\bf k})\cdot{\bm \tau}=
h_x({\bf k})\tau_x+h_y({\bf k})\tau_y+h_z({\bf k})\tau_z,
\end{equation}
with 
$h_x({\bf k})=-t_1\sum_j\cos({\bf k}\cdot{\bf e}_j)$, $h_y({\bf k})=-t_1\sum_j\sin({\bf k}\cdot{\bf e}_j)$ and $h_z({\bf k})=M-2t_2\sin\phi\sum_j\sin({\bf k}\cdot{\bf v}_j)$.
Here we have removed the trivial identity matrix term, with ${\bf e}_1=(0,a_0)$, ${\bf e}_2=(-\frac{\sqrt{3}a_0}{2},-\frac{a_0}{2})$, ${\bf e}_3=(\frac{\sqrt{3}a_0}{2},-\frac{a_0}{2})$
and ${\bf v}_1=(\sqrt{3}a_0,0)$, ${\bf v}_2=(-\frac{\sqrt{3}a_0}{2},\frac{3a_0}{2})$, ${\bf v}_3=-{\bf v}_2-{\bf v}_1$ ($a_0$ is the lattice constant). Moreover, we set the energy
difference between the two sublattices $M/t_1=-0.5$ if it is considered. Thus, with $\phi=\pi/2$ and $t_2=0.3t_1$, the noninteracting system lies in the topological phase with Chern number $C=1$~\cite{Haldane1988_S}.
The two energy bands read
\begin{equation}
{\cal E}_{\pm}({\bf k})=\pm\sqrt{h_x^2({\bf k})+h_y^2({\bf k})+h_z^2({\bf k})}\equiv \pm E_0({\bf k}).
\end{equation}

We write $H_0=\sum_{{\bf k},s=\pm\sigma}{\cal E}_s({\bf k})c_{{\bf k},s\sigma}^\dagger c_{{\bf k},s\sigma}$, with
\begin{align}
a_{{\bf k}\sigma}=\chi_+({\bf k})c_{{\bf k},+\sigma}+\chi_-({\bf k})c_{{\bf k},-\sigma},\quad b_{{\bf k}\sigma}=\xi_+({\bf k})c_{{\bf k},+\sigma}+\xi_-({\bf k})c_{{\bf k},-\sigma}.
\end{align}
One can easily obtain $\chi_+=\sqrt{\frac{1}{2E_0(E_0-h_z)}}(h_x-\ui h_y)$, $\chi_-=-\sqrt{\frac{1}{2E_0(E_0+h_z)}}(h_x-\ui h_y)$,
$\xi_+=\sqrt{\frac{E_0-h_z}{2E_0}}$ and $\xi_-=\sqrt{\frac{E_0+h_z}{2E_0}}$.
We further have
\begin{align}
a_{\bf p' \uparrow}^\dagger a_{\bf p \uparrow}a_{\bf q' \downarrow}^\dagger a_{\bf q\downarrow}=&\left[\chi_+^*({\bf p'}) c_{{\bf p'},+\uparrow}^\dagger+\chi_-^*({\bf p'}) c_{{\bf p}',-\uparrow}^\dagger\right]\left[\chi_+({\bf p}) c_{{\bf p},+\uparrow}+\chi_-({\bf p}) c_{{\bf p},-\uparrow}\right]\nonumber\\
&\left[\chi_+^*({\bf q'}) c_{{\bf q'},+\downarrow}^\dagger+\chi_-^*({\bf q'}) c_{{\bf q'},-\downarrow}^\dagger\right]\left[\chi_+({\bf q}) c_{{\bf q},+\downarrow}+\chi_-({\bf q}) c_{{\bf q},-\downarrow}\right] \nonumber\\
=&\sum_{s_1s_2s_3s_4} \chi_{s_1}^*({\bf p'})\chi_{s_2}({\bf p})\chi_{s_3}^*({\bf q'})\chi_{s_4}({\bf q})c_{{\bf p'},s_1\uparrow}^\dagger c_{{\bf p},s_2\uparrow}
c_{{\bf q'},s_3\downarrow}^\dagger c_{{\bf q},s_4\downarrow}
\end{align}
and, similarly,
\begin{align}
b_{\bf p' \uparrow}^\dagger b_{\bf p \uparrow}b_{\bf q' \downarrow}^\dagger b_{\bf q \downarrow}
=\sum_{s_1s_2s_3s_4} \xi_{s_1}^*({\bf p'})\xi_{s_2}({\bf p})\xi_{s_3}^*({\bf q'})\xi_{s_4}({\bf q})c_{{\bf p'},s_1\uparrow}^\dagger c_{{\bf p},s_2\uparrow}
c_{{\bf q'},s_3\downarrow}^\dagger c_{{\bf q},s_4\downarrow}
\end{align}
Thus, the on-site interaction
\begin{align}
H_I=U\sum_{\bf p'pq'q}\delta_{\bf p'+q'}^{\bf p+q}(a_{\bf p' \uparrow}^\dagger a_{\bf p \uparrow}a_{\bf q' \downarrow}^\dagger a_{\bf q\downarrow}
+b_{\bf p' \uparrow}^\dagger b_{\bf p \uparrow}b_{\bf q' \downarrow}^\dagger b_{\bf q \downarrow})
=\sum_{\substack{{\bf p'pq'q}\\ s_1s_2s_3s_4}}U_{\bf p'pq'q}^{s_1s_2s_3s_4}c_{{\bf p'},s_1\uparrow}^\dagger c_{{\bf p},s_2\uparrow}c_{{\bf q'},s_3\downarrow}^\dagger c_{{\bf q},s_4\downarrow},
\end{align}
where $U_{\bf p'pq'q}^{s_1s_2s_3s_4}\equiv U\delta^{\bf p+q}_{\bf p'+q'}\Lambda_{\bf p'pq'q}^{s_1s_2s_3s_4}$ and
\begin{equation}\label{Lambda}
\Lambda_{\bf p'pq'q}^{s_1s_2s_3s_4}\overset{\rm def}{=}\chi_{s_1}^*({\bf p'})\chi_{s_2}({\bf p})\chi_{s_3}^*({\bf q'})\chi_{s_4}({\bf q})+\xi_{s_1}^*({\bf p'})\xi_{s_2}({\bf p})\xi_{s_3}^*({\bf q'})\xi_{s_4}({\bf q}).
\end{equation}

For large $U$, we consider the symmetry-breaking order in $z$ direction, and write the Hubbard interaction in the mean-field form
\begin{align}
H_I=U\sum_i\left(\langle a^\dagger_{i\uparrow}a_{i\uparrow}\rangle a^\dagger_{i\downarrow}a_{i\downarrow}+a^\dagger_{i\uparrow}a_{i\uparrow} \langle a^\dagger_{i\downarrow}a_{i\downarrow}\rangle+
\langle b^\dagger_{i\uparrow}b_{i\uparrow}\rangle b^\dagger_{i\downarrow}b_{i\downarrow}+b^\dagger_{i\uparrow}b_{i\uparrow} \langle b^\dagger_{i\downarrow}b_{i\downarrow}\rangle\right)
\end{align}
Here $\langle\cdot\rangle$ is taken with respect to the mean-field ground state of the total Hamiltonian, which can be solved self-consistently.
We define the antiferromagnetic (AF) order $m_{\rm AF}\equiv \langle b^\dagger_{i\uparrow}b_{i\uparrow}-b^\dagger_{i\downarrow}b_{i\downarrow}\rangle U/2=-\langle a^\dagger_{i\uparrow}a_{i\uparrow}-a^\dagger_{i\downarrow}a_{i\downarrow}\rangle U/2$
and the charge order $m_{\rm C}=\langle a^\dagger_{i\uparrow}a_{i\uparrow}+a^\dagger_{i\downarrow}a_{i\downarrow}- b^\dagger_{i\uparrow}b_{i\uparrow}- b^\dagger_{i\downarrow}b_{i\downarrow}\rangle U/4$. After the Fourier transform, we have the Bloch Hamiltonian
\begin{align}
{\cal H}_{\rm MF}=
\left( \begin{array}{cc}
{\cal H}({\bf k})+m_\uparrow\tau_z   & \\
& {\cal H}({\bf k})+m_\downarrow\tau_z
\end{array} \right),
\end{align}
where the magnetic order $m_\uparrow=m_{\rm C}-m_{\rm AF}$ and $m_\downarrow=m_{\rm C}+m_{\rm AF}$.
Regarding the orders $m_{\rm C}$ and $m_{\rm AF}$ as the input parameters,
the Hamiltonian can be diagonalized as ${\cal H}_{\rm MF}({\bf k})=\sum_{s=\pm,\sigma=\uparrow\downarrow}\bar{{\cal E}}_{s\sigma}({\bf k})
\bar{c}_{{\bf k},s\sigma}^{\,\dagger}\bar{c}_{{\bf k},s\sigma}$, where
$\bar{{\cal E}}_{\pm\sigma}=\pm E_0^\sigma$,
with $E_0^\sigma\equiv\sqrt{E_0^2+2m_\sigma h_z+m_\sigma^2}$.
One can find the relation between the noninteracting and mean-field solutions ($\sigma=\uparrow\downarrow$):
\begin{align}
c_{{\bf k},+\sigma}=f_{++}^{\sigma}({\bf k})\bar{c}_{{\bf k},+\sigma}+f_{+-}^{\sigma}({\bf k})\bar{c}_{{\bf k},-\sigma},\quad
c_{{\bf k},-\sigma}=f_{-+}^{\sigma}({\bf k})\bar{c}_{{\bf k},+\sigma}+f_{--}^{\sigma}({\bf k})\bar{c}_{{\bf k},-\sigma},
\end{align}
with
\begin{align}
&|f_{++}^{\sigma}|^2+|f_{+-}^{\sigma}|^2=|f_{-+}^{\sigma}|^2+|f_{--}^{\sigma}|^2=1, \nonumber\\
&|f_{++}^{\sigma}|^2=|f_{--}^{\sigma}|^2,\quad |f_{+-}^{\sigma}|^2=|f_{-+}^{\sigma}|^2,\nonumber\\
&f_{++}^{\sigma\,*}f_{-+}^{\sigma}+f_{+-}^{\sigma\,*}f_{--}^{\sigma}=0.
\end{align}
The AF state at half-filling can be denoted as
 $|\Psi_{\rm MF}\rangle=\prod_{{\bf k}}\bar{c}_{{\bf k},-\uparrow}^{\,\dagger}\bar{c}_{{\bf k},-\downarrow}^{\,\dagger}|0\rangle$, and  $|\Psi_{\rm MF}\rangle\simeq\prod_{{\bf k}}a_{{\bf k}\uparrow}^{\,\dagger}b_{{\bf k}\downarrow}^{\,\dagger}|0\rangle$ when $m_{\rm AF}\to\infty$.

\section{II. Flow equations}

We study the interacting Haldane model by flow equation method.
The expansion parameter is the (small) interaction $U$ and normal ordering
is with respect to the AF state $|\Psi_{\rm MF}\rangle$, with
\begin{align}\label{correlation_MF}
\langle c_{{\bf k},s_1\sigma}^\dagger c_{{\bf k},s_2\sigma}\rangle=f_{s_1-}^{\sigma\,*}({\bf k})f_{s_2-}^{\sigma}({\bf k}), \quad
\langle c_{{\bf k},s_1\sigma} c_{{\bf k},s_2\sigma}^\dagger\rangle=f_{s_1+}^{\sigma}({\bf k})f_{s_2+}^{\sigma\,*}({\bf k}).
\end{align}
We start with the ansatz
\begin{align}
H(l)&=\sum_{\substack{{\bf k},s=\pm,\\\sigma=\uparrow\downarrow}}{\cal E}_s({\bf k}):c_{{\bf k},s\sigma}^\dagger c_{{\bf k},s\sigma}:
+\sum_{\substack{{\bf p'pq'q}\\ s_1s_2s_3s_4}}U_{\bf p'pq'q}^{s_1s_2s_3s_4}(l):c_{{\bf p'},s_1\uparrow}^\dagger c_{{\bf p},s_2\uparrow}
c_{{\bf q'},s_3\downarrow}^\dagger c_{{\bf q},s_4\downarrow}:,
\end{align}
where the interaction $U_{\bf p'pq'q}^{s_1s_2s_3s_4}(l)$ is responsible for the flow of the Hamiltonian and the flow of band energies and higher order terms
are neglected.
Since
\begin{align}
[:c_{{\bf k},s\sigma}^\dagger c_{{\bf k},s\sigma}:,\,:c_{{\bf p'},s_1\uparrow}^\dagger c_{{\bf p},s_2\uparrow}
c_{{\bf q'},s_3\downarrow}^\dagger c_{{\bf q},s_4\downarrow}:]
=\left(-\delta_{s}^{s_2}\delta_{\sigma}^\uparrow\delta_{\bf k}^{\bf p}-\delta_s^{s_4}\delta_{\sigma}^\downarrow\delta_{\bf k}^{\bf q}+\delta_s^{s_1}\delta_{\sigma}^\uparrow\delta_{\bf k}^{\bf p'}+\delta_s^{s_3}\delta_{\sigma}^\downarrow\delta_{\bf k}^{\bf q'}\right):c_{{\bf p'},s_1\uparrow}^\dagger c_{{\bf p},s_2\uparrow}c_{{\bf q'},s_3\downarrow}^\dagger c_{{\bf q},s_4\downarrow}:
\end{align}
we have the generator
\begin{align}\label{generator}
\eta(l)&=[H_0(l),H_I(l)]=\sum_{\substack{{\bf p'pq'q}\\ s_1s_2s_3s_4}}U_{\bf p'pq'q}^{s_1s_2s_3s_4}(l)\Delta_{\bf p'pq'q}^{s_1s_2s_3s_4}:c_{{\bf p'},s_1\uparrow}^\dagger c_{{\bf p},s_2\uparrow}
c_{{\bf q'},s_3\downarrow}^\dagger c_{{\bf q},s_4\downarrow}:,
\end{align}
where $\Delta_{\bf p'pq'q}^{s_1s_2s_3s_4}\overset{\rm def}{=}{\cal E}_{s_1}({\bf p}')-{\cal E}_{s_2}({\bf p})+{\cal E}_{s_3}({\bf q}')-{\cal E}_{s_4}({\bf q})$ is the energy difference
before and after scattering.
Since
\begin{align}
[\eta(l),H_0(l)]
=-\sum_{\substack{{\bf p'pq'q}\\ s_1s_2s_3s_4}}U_{\bf p'pq'q}^{s_1s_2s_3s_4}(l)(\Delta_{\bf p'pq'q}^{s_1s_2s_3s_4})^2
:c_{{\bf p'},s_1\uparrow}^\dagger c_{{\bf p},s_2\uparrow}
c_{{\bf q'},s_3\downarrow}^\dagger c_{{\bf q},s_4\downarrow}:,
\end{align}
the flow of the interaction is given by
\begin{equation}
U_{\bf p'pq'q}^{s_1s_2s_3s_4}(l)=U\delta^{\bf p+q}_{\bf p'+q'}\Lambda_{\bf p'pq'q}^{s_1s_2s_3s_4}\exp[-l(\Delta_{\bf p'pq'q}^{s_1s_2s_3s_4})^2],
\end{equation}
which decays to zero when the flow parameter $l\to\infty$.

Next we work out the flow equation transformation for the creation operators. Since $a^\dagger_{{\bf k}\sigma}=\chi^*_+ c^\dagger_{{\bf k},+\sigma}+\chi^*_- c^\dagger_{{\bf k},-\sigma}$, $b^\dagger_{{\bf k}\sigma}=\xi^*_+ c^\dagger_{{\bf k},+\sigma}
+\xi^*_- c_{{\bf k},-\sigma}$ and the relations
\begin{align}
[:c_{{\bf p'},s_1\uparrow}^\dagger c_{{\bf p},s_2\uparrow}c_{{\bf q'},s_3\downarrow}^\dagger c_{{\bf q},s_4\downarrow}:,\, c_{{\bf k},{s\sigma}}^\dagger]
=\delta_{s}^{s_2}\delta_{\sigma}^\uparrow\delta_{\bf k}^{\bf p}:c_{{\bf p'},s_1\uparrow}^\dagger c_{{\bf q'},s_3\downarrow}^\dagger c_{{\bf q},s_4\downarrow}:
+\delta_{s}^{s_4}\delta_{\sigma}^\downarrow\delta_{\bf k}^{\bf q}:c_{{\bf p'},s_1\uparrow}^\dagger c_{{\bf p},s_2\uparrow}
c_{{\bf q'},s_3\downarrow}^\dagger:,
\end{align}
we assume
\begin{align}\label{Al_Flow}
{\cal A}_{{\bf k}\uparrow}^\dagger(l)&=h_{{\bf k},+}(l)c_{{\bf k},{+\uparrow}}^\dagger+h_{{\bf k},-}(l)c_{{\bf k},{-\uparrow}}^\dagger
+\sum_{\substack{{\bf p'q'q}\\ \mu\nu\gamma}}M_{{\bf k},{\bf p'q'q}}^{\mu\nu\gamma}(l)\delta_{\bf p'+q'}^{\bf k+q} :c_{{\bf p'},\mu\uparrow}^\dagger c_{{\bf q'},\nu\downarrow}^\dagger c_{{\bf q},\gamma\downarrow}:, \nonumber\\
{\cal A}_{{\bf k}\downarrow}^\dagger(l)&=g_{{\bf k},+}(l)c_{{\bf k},{+\downarrow}}^\dagger+g_{{\bf k},-}(l)c_{{\bf k},{-\downarrow}}^\dagger
+\sum_{\substack{{\bf p'pq'}\\ \mu\nu\gamma}}W_{{\bf k},{\bf p'pq'}}^{\mu\nu\gamma}(l)\delta_{\bf p'+q'}^{\bf p+k} :c_{{\bf p'},\mu\uparrow}^\dagger c_{{\bf p},\nu\uparrow}
c_{{\bf q'},\gamma\downarrow}^\dagger:.
\end{align}
Here $h_{{\bf k},+}(l=0)=g_{{\bf k},+}(l=0)=\chi^*_+({\bf k})$, $h_{{\bf k},-}(l=0)=g_{{\bf k},-}(l=0)=\chi^*_-({\bf k})$, and $M_{{\bf k},{\bf p'q'q}}^{\mu\nu\gamma}(l=0)=W_{{\bf k},{\bf p'pq'}}^{\mu\nu\gamma}(l=0)=0$. The operators ${\cal B}_{{\bf k}\sigma}^\dagger(l)$ take the same form as in Eq.~(\ref{Al_Flow}) but with
$h_{{\bf k},+}(l=0)=g_{{\bf k},+}(l=0)=\xi^*_+({\bf k})$ and $h_{{\bf k},-}(l=0)=g_{{\bf k},-}(l=0)=\xi^*_-({\bf k})$.
With
\begin{align}
&[:c_{{\bf p'},s_1\uparrow}^\dagger c_{{\bf p},s_2\uparrow}c_{{\bf q'},s_3\downarrow}^\dagger c_{{\bf q},s_4\downarrow}:,\,  :c_{{\bf 1'},\mu\uparrow}^\dagger c_{{\bf 2'},\nu\downarrow}^\dagger c_{{\bf 2},\gamma\downarrow}:]\nonumber\\
=&-\delta_{s_3}^\gamma\delta_{\bf q'}^{\bf 2}:c_{{\bf p'},s_1\uparrow}^\dagger c_{{\bf p},s_2\uparrow}c_{{\bf q},s_4\downarrow}c_{{\bf 1'},\mu\uparrow}^\dagger c_{{\bf 2'},\nu\downarrow}^\dagger:+\delta_{s_2}^\mu\delta_{\bf p}^{\bf 1'}:c_{{\bf p'},s_1\uparrow}^\dagger c_{{\bf q'},s_3\downarrow}^\dagger c_{{\bf q},s_4\downarrow}c_{{\bf 2'},\nu\downarrow}^\dagger c_{{\bf 2},\gamma\downarrow}:\nonumber\\
&-\delta_{s_4}^\nu\delta_{\bf q}^{\bf 2'}:c_{{\bf p'},s_1\uparrow}^\dagger c_{{\bf p},s_2\uparrow}c_{{\bf q'},s_3\downarrow}^\dagger c_{{\bf 1'},\mu\uparrow}^\dagger c_{{\bf 2},\gamma\downarrow}:
+(f_{s_3-}^{\downarrow\,*}f_{\gamma-}^\downarrow\delta_{\bf q'}^{\bf 2}f_{s_2+}^{\uparrow}f_{\mu+}^{\uparrow\,*}\delta_{\bf p}^{\bf 1'}-
f_{\mu-}^{\uparrow\,*}f_{s_2-}^{\uparrow}\delta_{\bf p}^{\bf 1'}f_{\gamma+}^{\downarrow}f_{s_3+}^{\downarrow\,*}\delta_{\bf q'}^{\bf 2})\nonumber \\
&:c_{{\bf p'},s_1\uparrow}^\dagger c_{{\bf q},s_4\downarrow}c_{{\bf 2'},\nu\downarrow}^\dagger:
+(f_{s_3-}^{\downarrow\,*}f_{\gamma-}^\downarrow\delta_{\bf q'}^{\bf 2}f_{s_4+}^{\downarrow}f_{\nu+}^{\downarrow\,*}\delta_{\bf q}^{\bf 2'}-
f_{\gamma+}^{\downarrow}f_{s_3+}^{\downarrow\,*}\delta_{\bf q'}^{\bf 2}f_{\nu-}^{\downarrow\,*}f_{s_4-}^\downarrow\delta_{\bf q}^{\bf 2'})
:c_{{\bf p'},s_1\uparrow}^\dagger c_{{\bf p},s_2\uparrow}c_{{\bf 1'},\mu\uparrow}^\dagger:\nonumber\\
&+(f_{s_2+}^{\uparrow}f_{\mu+}^{\uparrow\,*}\delta_{\bf p}^{\bf 1'}f_{s_4+}^{\downarrow}f_{\nu+}^{\downarrow\,*}\delta_{\bf q}^{\bf 2'}-
f_{\mu-}^{\uparrow\,*}f_{s_2-}^{\uparrow}\delta_{\bf p}^{\bf 1'}f_{\nu-}^{\downarrow\,*}f_{s_4-}^{\downarrow}\delta_{\bf q}^{\bf 2'} )
:c_{{\bf p'},s_1\uparrow}^\dagger c_{{\bf q'},s_3\downarrow}^\dagger c_{{\bf 2},\gamma\downarrow}: \nonumber\\
&+\delta_{\bf p}^{\bf 1'}\delta_{\bf q'}^{\bf 2}\delta_{\bf q}^{\bf 2'}(f_{s_2+}^{\uparrow}f_{\mu+}^{\uparrow\,*}f_{s_3-}^{\downarrow\,*}f_{\gamma-}^{\downarrow}f_{s_4+}^{\downarrow}f_{\nu+}^{\downarrow\,*}+f_{\mu-}^{\uparrow\,*}f_{s_2-}^{\uparrow}f_{\gamma+}^{\downarrow}f_{s_3+}^{\downarrow\,*}f_{\nu-}^{\downarrow\,*}f_{s_4-}^{\downarrow})c_{{\bf p'},s_1\uparrow}^\dagger,
\end{align}
and
\begin{align}
&[:c_{{\bf p'},s_1\uparrow}^\dagger c_{{\bf p},s_2\uparrow}c_{{\bf q'},s_3\downarrow}^\dagger c_{{\bf q},s_4\downarrow}:,\,  :c_{{\bf 1'},\mu\uparrow}^\dagger c_{{\bf 1},\nu\uparrow}c_{{\bf 2'},\gamma\downarrow}^\dagger :]\nonumber\\
=&\delta_{s_1}^\nu\delta_{\bf p'}^{\bf 1}:c_{{\bf p},s_2\uparrow}c_{{\bf q'},s_3\downarrow}^\dagger c_{{\bf q},s_4\downarrow}c_{{\bf 1'},\mu\uparrow}^\dagger c_{{\bf 2'},\gamma\downarrow}^\dagger : +
\delta_{s_2}^{\mu}\delta_{\bf p}^{\bf 1'}:c_{{\bf p'},s_1\uparrow}^\dagger c_{{\bf q'},s_3\downarrow}^\dagger c_{{\bf q},s_4\downarrow}c_{{\bf 1},\nu\uparrow}c_{{\bf 2'},\gamma\downarrow}^\dagger :\nonumber\\
&+\delta_{s_4}^{\gamma}\delta_{\bf q}^{\bf 2'}:c_{{\bf p'},s_1\uparrow}^\dagger c_{{\bf p},s_2\uparrow}c_{{\bf q'},s_3\downarrow}^\dagger c_{{\bf 1'},\mu\uparrow}^\dagger c_{{\bf 1},\nu\uparrow}:
+(f_{s_1-}^{\uparrow\,*}f_{\nu-}^{\uparrow}\delta_{\bf p'}^{\bf 1}f_{s_2+}^{\uparrow}f_{\mu+}^{\uparrow\,*}\delta_{\bf p}^{\bf 1'}-f_{\nu+}^{\uparrow}f_{s_1+}^{\uparrow\,*}\delta_{\bf p'}^{\bf 1}f_{\mu-}^{\uparrow\,*}f_{s_2-}^{\uparrow}\delta_{\bf p}^{\bf 1'}) \nonumber\\
&:c_{{\bf q'},s_3\downarrow}^\dagger c_{{\bf q},s_4\downarrow}c_{{\bf 2'},\gamma\downarrow}^\dagger:
+(-f_{s_1-}^{\uparrow\,*}f_{\nu-}^{\uparrow}\delta_{\bf p'}^{\bf 1}f_{s_4+}^{\downarrow}f_{\gamma+}^{\downarrow\,*}\delta_{\bf q}^{\bf 2'}+f_{\nu+}^{\uparrow}f_{s_1+}^{\uparrow\,*}\delta_{\bf p'}^{\bf 1}f_{\gamma-}^{\downarrow\,*}f_{s_4-}^{\downarrow}\delta_{\bf q}^{\bf 2'})
: c_{{\bf p},s_2\uparrow}c_{{\bf q'},s_3\downarrow}^\dagger c_{{\bf 1'},\mu\uparrow}^\dagger: \nonumber\\
&+(-f_{s_2+}^{\uparrow}f_{\mu+}^{\uparrow\,*}\delta_{\bf p}^{\bf 1'}f_{s_4+}^{\downarrow}f_{\gamma+}^{\downarrow\,*}\delta_{\bf q}^{\bf 2'}+f_{\mu-}^{\uparrow\,*}f_{s_2-}^{\uparrow}\delta_{\bf p}^{\bf 1'}f_{\gamma-}^{\downarrow\,*}f_{s_4-}^{\downarrow}\delta_{\bf q}^{\bf 2'})
:c_{{\bf p'},s_1\uparrow}^\dagger c_{{\bf q'},s_3\downarrow}^\dagger c_{{\bf 1},\nu\uparrow}: \nonumber\\
&+\delta_{\bf p'}^{\bf 1}\delta_{\bf p}^{\bf 1'}\delta_{\bf q}^{\bf 2'}(f_{s_1-}^{\uparrow\,*}f_{\nu-}^{\uparrow}f_{s_2+}^{\uparrow}f_{\mu+}^{\uparrow\,*}f_{s_4+}^{\downarrow}f_{\gamma+}^{\downarrow\,*}+f_{\nu+}^{\uparrow}f_{s_1+}^{\uparrow\,*}f_{\mu-}^{\uparrow\,*}f_{s_2-}^{\uparrow}f_{\gamma-}^{\downarrow\,*}f_{s_4-}^{\downarrow})c_{{\bf q'},s_3\downarrow}^\dagger,
\end{align}
we obtain the leading-order flow equations for the creation operators
\begin{align}\label{solu_hM}
&\frac{\partial h_{{\bf k},+}(l)}{\partial l}=\sum_{{\bf p'q'q}}\sum_{\substack{s_2s_3s_4\\ \mu\nu\gamma}}
F_{\bf p'qq'}^{s_2s_3s_4,\mu\gamma\nu}M_{{\bf k},{\bf p'q'q}}^{\mu\nu\gamma}(l)U_{\bf kp'qq'}^{+s_2s_3s_4}(l)\Delta_{\bf kp'qq'}^{+s_2s_3s_4},\nonumber\\
&\frac{\partial h_{{\bf k},-}(l)}{\partial l}=\sum_{{\bf p'q'q}}\sum_{\substack{s_2s_3s_4\\ \mu\nu\gamma}}
F_{\bf p'qq'}^{s_2s_3s_4,\mu\gamma\nu}M_{{\bf k},{\bf p'q'q}}^{\mu\nu\gamma}(l)U_{\bf kp'qq'}^{-s_2s_3s_4}(l)\Delta_{\bf kp'qq'}^{-s_2s_3s_4},\nonumber\\
&\frac{\partial M_{{\bf k},{\bf p'q'q}}^{\mu\nu\gamma}(l)}{\partial l}=h_{{\bf k},+}(l)\Delta_{\bf p'kq'q}^{\mu +\nu\gamma}U_{\bf p'kq'q}^{\mu +\nu\gamma}(l)
+h_{{\bf k},-}(l)\Delta_{\bf p'kq'q}^{\mu -\nu\gamma}U_{\bf p'kq'q}^{\mu -\nu\gamma}(l),
\end{align}
and
\begin{align}\label{solu_gW}
&\frac{\partial g_{{\bf k},+}(l)}{\partial l}=\sum_{{\bf p'pq'}}\sum_{\substack{s_1s_2s_4\\ \mu\nu\gamma}}
G_{\bf pp'q'}^{s_1s_2s_4,\nu\mu\gamma}W_{{\bf k},{\bf p'pq'}}^{\mu\nu\gamma}(l)U_{\bf pp'kq'}^{s_1s_2+s_4}(l)\Delta_{\bf pp'kq'}^{s_1s_2+s_4},\nonumber\\
&\frac{\partial g_{{\bf k},-}(l)}{\partial l}=\sum_{{\bf p'pq'}}\sum_{\substack{s_1s_2s_4\\ \mu\nu\gamma}}
G_{\bf pp'q'}^{s_1s_2s_4,\nu\mu\gamma}W_{{\bf k},{\bf p'pq'}}^{\mu\nu\gamma}(l)U_{\bf pp'kq'}^{s_1s_2-s_4}(l)\Delta_{\bf pp'kq'}^{s_1s_2-s_4},\nonumber\\
&\frac{\partial W_{{\bf k},{\bf p'pq'}}^{\mu\nu\gamma}(l)}{\partial l}=g_{{\bf k},+}(l)U_{\bf p'pq'k}^{\mu\nu\gamma+}(l)\Delta_{\bf p'pq'k}^{\mu\nu\gamma+}
+g_{{\bf k},-}(l)U_{\bf p'pq'k}^{\mu\nu\gamma-}(l)\Delta_{\bf p'pq'k}^{\mu\nu\gamma-},
\end{align}
where $F_{\bf p'qq'}^{s_2s_3s_4,\mu\gamma\nu}=f_{s_2+}^{\uparrow}({\bf p}')f_{\mu+}^{\uparrow\,*}({\bf p}')f_{s_3-}^{\downarrow\,*}({\bf q})f_{\gamma-}^{\downarrow}({\bf q})f_{s_4+}^{\downarrow}({\bf q}')f_{\nu+}^{\downarrow\,*}({\bf q}')+f_{s_2-}^{\uparrow}({\bf p}')f_{\mu-}^{\uparrow\,*}({\bf p}')f_{s_3+}^{\downarrow\,*}({\bf q})f_{\gamma+}^{\downarrow}({\bf q})f_{s_4-}^{\downarrow}({\bf q}')f_{\nu-}^{\downarrow\,*}({\bf q}')$ and
$G_{\bf pp'q'}^{s_1s_2s_4,\nu\mu\gamma}=f_{s_1-}^{\uparrow\,*}({\bf p})f_{\nu-}^{\uparrow}({\bf p})f_{s_2+}^{\uparrow}({\bf p}')f_{\mu+}^{\uparrow\,*}({\bf p}')f_{s_4+}^{\downarrow}({\bf q}')f_{\gamma+}^{\downarrow\,*}({\bf q}')+f_{s_1+}^{\uparrow\,*}({\bf p})f_{\nu+}^{\uparrow}({\bf p})f_{s_2-}^{\uparrow}({\bf p}')f_{\mu-}^{\uparrow\,*}({\bf p}')f_{s_4-}^{\downarrow}({\bf q}')f_{\gamma-}^{\downarrow\,*}({\bf q}')$.

We adopt the forward-backward transformation~\cite{Moeckel2008_S,Moeckel2009_S} to calculate the time-evolved operators. The forward (or backward)
transformations are derived by integrating the flow equations (\ref{solu_hM}) and (\ref{solu_gW}) from $l=0$ to $\infty$ (or from $l=\infty$ to 0)
with different initial conditions. We keep the terms up to second order in $U$ and obtain the approximate analytic solutions.
Take the number operator ${\cal N}^A_{{\bf k}\uparrow}(l)={\cal A}_{{\bf k}\uparrow}^\dagger(l){\cal A}_{{\bf k}\uparrow}(l)$ as an example.
Time evolution yields
\begin{align}\label{evolution}
&h_{{\bf k},+}(l=\infty,t)=h_{{\bf k},+}(l=\infty,t=0)e^{-\ui{\cal E}_+({\bf k})t},\nonumber\\
&h_{{\bf k},-}(l=\infty,t)=h_{{\bf k},-}(l=\infty,t=0)e^{-\ui{\cal E}_-({\bf k})t},\nonumber\\
&M_{{\bf k},{\bf p'q'q}}^{\mu\nu\gamma}(l=\infty,t)=M_{{\bf k},{\bf p'q'q}}^{\mu\nu\gamma}(l=\infty,t=0)
e^{-\ui[{\cal E}_\mu({\bf p'})+{\cal E}_\nu({\bf q'})-{\cal E}_\gamma({\bf q})]t}.
\end{align}
Since
\begin{align}
\langle:c_{{\bf p'},\mu\uparrow}^\dagger c_{{\bf q'},\nu\downarrow}^\dagger c_{{\bf q},\gamma\downarrow}:\, :c_{{\bf 2},s_3\downarrow}^\dagger c_{{\bf 2'},s_2\downarrow} c_{{\bf 1'},s_1\uparrow}:\rangle
&=\delta_{\bf p'}^{\bf 1'}f_{\mu-}^{\uparrow\,*}({\bf p}')f_{s_1-}^{\uparrow}({\bf p}')\delta_{\bf q'}^{\bf 2'}f_{\nu-}^{\downarrow\,*}({\bf q'})f_{s_2-}^{\downarrow}({\bf q'})\delta_{\bf q}^{\bf 2}f_{\gamma+}^{\downarrow}({\bf q})f_{s_3+}^{\downarrow\,*}({\bf q}),
\end{align}
we obtain the distribution of spin-up particles at $A$ sites
\begin{align}
&N^A_{{\bf k}\uparrow}(t)\overset{\rm def}{=}\langle\Psi|{\cal N}^A_{{\bf k}\uparrow}(l=0,t)|\Psi\rangle\nonumber\\
&=|h_{{\bf k},+}(0,t)|^2|f_{+-}^\uparrow({\bf k})|^2+|h_{{\bf k},-}(0,t)|^2|f_{--}^\uparrow({\bf k})|^2
+2\Re\left[h_{{\bf k},+}(0,t)h_{{\bf k},-}^*(0,t)f_{+-}^{\uparrow\,*}({\bf k})f_{--}^{\uparrow}({\bf k})\right]\nonumber\\
&+\sum_{{\bf p'q'q}}\sum_{\substack{s_1s_2s_3\\ \mu\nu\gamma}}\delta_{\bf p'+q'}^{\bf k+q}M_{{\bf k},{\bf p'q'q}}^{s_1s_2s_3\,*}(0,t)
M_{{\bf k},{\bf p'q'q}}^{\mu\nu\gamma}(0,t)f_{\mu-}^{\uparrow\,*}({\bf p}')f_{s_1-}^{\uparrow}({\bf p}')f_{\nu-}^{\downarrow\,*}({\bf q'})f_{s_2-}^{\downarrow}({\bf q'})f_{\gamma+}^{\downarrow}({\bf q})f_{s_3+}^{\downarrow\,*}({\bf q}).
\end{align}
The computation of $h_{{\bf k},\pm}(l=0,t)$ and $M_{{\bf k},{\bf p'q'q}}^{\mu\nu\gamma}(l=0,t)$ is achieved by composing the forward transformation (FT),
the time evolution (TE) and the backward transformation (BT), such as
\begin{align}
h_{{\bf k},\pm}(l=0,t=0)\xrightarrow{\rm FT}h_{{\bf k},\pm}(l=\infty,t=0)\xrightarrow{\rm TE}h_{{\bf k},\pm}(l=\infty,t)\xrightarrow{\rm BT}h_{{\bf k},\pm}(l=0,t).
\end{align}

Up to now, the analytic solutions are very complicated despite the neglect of higher order terms.
It is mainly due to the various possible scattering channels in the flow equations (\ref{solu_hM}) and (\ref{solu_gW}).
To simplify the analysis, we take into account only the major contribution, i.e.
\begin{align}\label{FG_simplified}
F_{\bf p'qq'}^{s_2s_3s_4,\mu\gamma\nu}&\simeq\delta_{s_2}^{\mu}\delta_{s_3}^{\gamma}\delta_{s_4}^{\nu}{\cal F}_{\bf p'qq'}^{\mu\gamma\nu},\nonumber\\
G_{\bf pp'q'}^{s_1s_2s_4,\nu\mu\gamma}&\simeq\delta_{s_1}^{\nu}\delta_{s_2}^{\mu}\delta_{s_4}^{\gamma}{\cal G}_{\bf pp'q'}^{\nu\mu\gamma},
\end{align}
where ${\cal F}_{\bf p'qq'}^{\mu\gamma\nu}\overset{\rm def}{=}n_{\mu+}^{\uparrow}({\bf p}')n_{\gamma-}^{\downarrow}({\bf q})n_{\nu+}^{\downarrow}({\bf q}')+n_{\mu-}^{\uparrow}({\bf p}')n_{\gamma+}^{\downarrow}({\bf q})n_{\nu-}^{\downarrow}({\bf q}')$ and ${\cal G}_{\bf pp'q'}^{\nu\mu\gamma}\overset{\rm def}{=}n_{\nu-}^{\uparrow}({\bf p})n_{\mu+}^{\uparrow}({\bf p}')n_{\gamma+}^{\downarrow}({\bf q}')+n_{\nu+}^{\uparrow}({\bf p})n_{\mu-}^{\uparrow}({\bf p}')n_{\gamma-}^{\downarrow}({\bf q}')$ with $n_{s_1s_2}^{\sigma}({\bf k})\equiv|f_{s_1s_2}^{\sigma}({\bf k})|^2$.

\section{III. Pseudospin dynamics}

\begin{figure}
\includegraphics[width=0.58\textwidth]{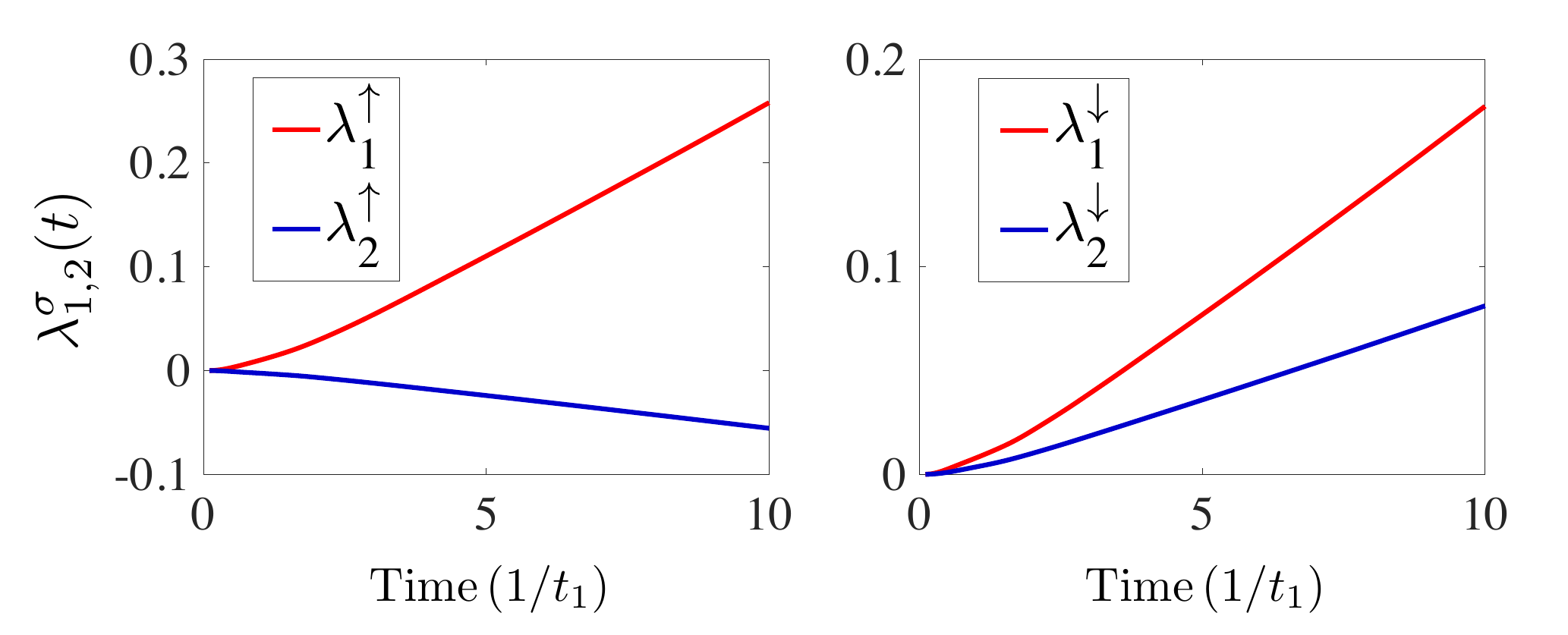}
\caption{Time dependence of the parameters $\lambda^{\sigma}_{1,2}(t)$. In a short time, they are all approximately linear in $t$. Here the values are taken at
${\bf k}=2({\bf b}_1+{\bf b}_2)/5$.
}\label{figS1}
\end{figure}

For convenience's sake, we denote ${\cal E}_+({\bf k})-{\cal E}_-({\bf k})=2E_0({\bf k})\equiv 1/T_{\rm g}({\bf k})$, $X_{\sigma}({\bf k},t)\equiv\langle\tau_{x}^\sigma({\bf k},t)\rangle$, $Y_{\sigma}({\bf k},t)\equiv\langle\tau_{y}^\sigma({\bf k},t)\rangle$ and $Z_{\sigma}({\bf k},t)\equiv\langle\tau_{z}^\sigma({\bf k},t)\rangle$ in the following.
By the forward-backward transformation, we have the results
\begin{align}\label{XYZ}
X_\sigma({\bf k},t)&=X_\sigma^{(0)}({\bf k})+X_\sigma^{(c)}({\bf k},t)+X_\sigma^{(h)}({\bf k},t)+X_\sigma^{(l)}({\bf k},t),\nonumber\\
Y_\sigma({\bf k},t)&=Y_\sigma^{(0)}({\bf k})+Y_\sigma^{(c)}({\bf k},t)+Y_\sigma^{(h)}({\bf k},t)+Y_\sigma^{(l)}({\bf k},t),\nonumber\\
Z_\sigma({\bf k},t)&=Z_\sigma^{(0)}({\bf k})+Z_\sigma^{(c)}({\bf k},t)+Z_\sigma^{(h)}({\bf k},t)+Z_\sigma^{(l)}({\bf k},t),
\end{align}
where the incoherent part
\begin{align}\label{incoherent}
X_\sigma^{(0)}({\bf k})&=\frac{h_x}{E_0}[n_{+-}^\sigma({\bf k})-n_{--}^\sigma({\bf k})],\nonumber\\
Y_\sigma^{(0)}({\bf k})&=\frac{h_y}{E_0}[n_{+-}^\sigma({\bf k})-n_{--}^\sigma({\bf k})],\nonumber\\
Z_\sigma^{(0)}({\bf k})&=\frac{h_z}{E_0}[n_{+-}^\sigma({\bf k})-n_{--}^\sigma({\bf k})],
\end{align}
the coherent time-dependent oscillation
\begin{align}\label{coherent}
X_{\sigma}^{(c)}({\bf k},t)=&\frac{h_z}{E_0}\frac{2h_x}{\sqrt{E_0^2-h_z^2}}f^{\sigma\,*}_{--}({\bf k})f^{\sigma}_{+-}({\bf k})\cos\left(t/T_{\rm g}\right)+\frac{2h_y}{\sqrt{E_0^2-h_z^2}}f^{\sigma\,*}_{--}({\bf k})f^{\sigma}_{+-}({\bf k})\sin\left(t/T_{\rm g}\right),\nonumber\\
Y_{\sigma}^{(c)}({\bf k},t)=&\frac{h_z}{E_0}\frac{2h_y}{\sqrt{E_0^2-h_z^2}}f^{\sigma\,*}_{--}({\bf k})f^{\sigma}_{+-}({\bf k})\cos\left(t/T_{\rm g}\right)-\frac{2h_x}{\sqrt{E_0^2-h_z^2}}f^{\sigma\,*}_{--}({\bf k})f^{\sigma}_{+-}({\bf k})\sin\left(t/T_{\rm g}\right),\nonumber\\
Z_{\sigma}^{(c)}({\bf k},t)=&-\frac{2}{E_0}\sqrt{E_0^2-h_z^2}f^{\sigma\,*}_{--}({\bf k})f^{\sigma}_{+-}({\bf k})\cos\left(t/T_{\rm g}\right),
\end{align}
the interaction-induced high-frequency fluctuation
\begin{align}\label{high_fuctuation}
X_\sigma^{(h)}({\bf k},t)&\approx-2\lambda_1^\sigma(t)\left[\frac{h_xh_z}{E_0\sqrt{E_0^2-h_z^2}}f^{\sigma\,*}_{--}({\bf k})f^{\sigma}_{+-}({\bf k})\cos\left(t/T_{\rm g}\right)+\frac{h_y}{\sqrt{E_0^2-h_z^2}}f^{\sigma\,*}_{--}({\bf k})f^{\sigma}_{+-}({\bf k})\sin\left(t/T_{\rm g}\right)\right],\nonumber\\
Y_\sigma^{(h)}({\bf k},t)&\approx-2\lambda_1^\sigma(t)\left[\frac{h_yh_z}{E_0\sqrt{E_0^2-h_z^2}}f^{\sigma\,*}_{--}({\bf k})f^{\sigma}_{+-}({\bf k})\cos\left(t/T_{\rm g}\right)-\frac{h_x}{\sqrt{E_0^2-h_z^2}}f^{\sigma\,*}_{--}({\bf k})f^{\sigma}_{+-}({\bf k})\sin\left(t/T_{\rm g}\right)\right],\nonumber\\
Z_\sigma^{(h)}({\bf k},t)&\approx2\lambda_1^\sigma(t)\frac{\sqrt{E_0^2-h_z^2}}{E_0}f^{\sigma\,*}_{--}({\bf k})f^{\sigma}_{+-}({\bf k})\cos\left(t/T_{\rm g}\right),
\end{align}
with
\begin{align}\label{lambda1}
\lambda_1^\uparrow(t)\overset{\rm def}{=}&2U^2\sum_{\bf p'q'q}\sum_{\mu\nu\gamma}\delta_{\bf p'+q'}^{\bf k+q}{\cal F}_{\bf p'qq'}^{\mu\gamma\nu}
\left[\frac{|\Lambda_{\bf kp'qq'}^{-\mu\gamma\nu}|^2}{(\Delta_{\bf kp'qq'}^{-\mu\gamma\nu})^2}\sin^2\left(\frac{\Delta_{\bf kp'qq'}^{-\mu\gamma\nu}\,t}{2}\right)+\frac{|\Lambda_{\bf kp'qq'}^{+\mu\gamma\nu}|^2}{(\Delta_{\bf kp'qq'}^{+\mu\gamma\nu})^2}\sin^2\left(\frac{\Delta_{\bf kp'qq'}^{+\mu\gamma\nu}\,t}{2}\right)\right], \nonumber\\
\lambda_1^\downarrow(t)\overset{\rm def}{=}&2U^2\sum_{\bf p'pq'}\sum_{\mu\nu\gamma}\delta_{\bf p'+q'}^{\bf p+k}{\cal G}_{\bf pp'q'}^{\nu\mu\gamma}
\left[\frac{|\Lambda_{\bf pp'kq'}^{\nu\mu-\gamma}|^2}{(\Delta_{\bf pp'kq'}^{\nu\mu-\gamma})^2}\sin^2\left(\frac{\Delta_{\bf pp'kq'}^{\nu\mu-\gamma}\,t}{2}\right)+\frac{|\Lambda_{\bf pp'kq'}^{\nu\mu+\gamma}|^2}{(\Delta_{\bf pp'kq'}^{\nu\mu+\gamma})^2}\sin^2\left(\frac{\Delta_{\bf pp'kq'}^{\nu\mu+\gamma}\,t}{2}\right)\right],
\end{align}
and the interaction-induced low-frequency fluctuation
\begin{align}\label{low_fuctuation}
X_\sigma^{(l)}({\bf k},t)\approx-2\lambda_2^\sigma(t)\frac{h_x}{E_0},\quad\quad
Y_\sigma^{(l)}({\bf k},t)\approx-2\lambda_2^\sigma(t)\frac{h_y}{E_0},\quad\quad
Z_\sigma^{(l)}({\bf k},t)\approx-2\lambda_2^\sigma(t)\frac{h_z}{E_0},
\end{align}
with
\begin{align}\label{lambda2}
\lambda_2^\uparrow(t)\overset{\rm def}{=}&2U^2\sum_{\bf p'q'q}\sum_{\mu\nu\gamma}\delta_{\bf p'+q'}^{\bf k+q}\left[\frac{{\cal I}_{\bf kp'qq'}^{+\mu\gamma\nu}|\Lambda_{\bf kp'qq'}^{+\mu\gamma\nu}|^2}{(\Delta_{\bf kp'qq'}^{+\mu\gamma\nu})^2}\sin^2\left(\frac{\Delta_{\bf kp'qq'}^{+\mu\gamma\nu}\,t}{2}\right)-
\frac{{\cal I}_{\bf kp'qq'}^{-\mu\gamma\nu}|\Lambda_{\bf kp'qq'}^{-\mu\gamma\nu}|^2}{(\Delta_{\bf kp'qq'}^{-\mu\gamma\nu})^2}\sin^2\left(\frac{\Delta_{\bf kp'qq'}^{-\mu\gamma\nu}\,t}{2}\right)\right], \nonumber\\
\lambda_2^\downarrow(t)\overset{\rm def}{=}&2U^2\sum_{\bf p'pq'}\sum_{\mu\nu\gamma}\delta_{\bf p'+q'}^{\bf p+k}\left[\frac{{\cal J}_{\bf pp'kq'}^{\nu\mu+\gamma}|\Lambda_{\bf pp'kq'}^{\nu\mu+\gamma}|^2}{(\Delta_{\bf pp'kq'}^{\nu\mu+\gamma})^2}\sin^2\left(\frac{\Delta_{\bf pp'kq'}^{\nu\mu+\gamma}\,t}{2}\right)-
\frac{{\cal J}_{\bf pp'kq'}^{\nu\mu-\gamma}|\Lambda_{\bf pp'kq'}^{\nu\mu-\gamma}|^2}{(\Delta_{\bf pp'kq'}^{\nu\mu-\gamma})^2}\sin^2\left(\frac{\Delta_{\bf pp'kq'}^{\nu\mu-\gamma}\,t}{2}\right)\right].
\end{align}
Here we have denoted ($s=\pm$)
\begin{align}\label{IJ_MF}
{\cal I}_{\bf kp'qq'}^{s\mu\gamma\nu}&\overset{\rm def}{=}n_{s-}^\uparrow({\bf k})n_{\mu+}^{\uparrow}({\bf p}')n_{\gamma-}^{\downarrow}({\bf q})n_{\nu+}^{\downarrow}({\bf q}')-n_{s+}^\uparrow({\bf k})n_{\mu-}^{\uparrow}({\bf p}')n_{\gamma+}^{\downarrow}({\bf q})n_{\nu-}^{\downarrow}({\bf q}'),\nonumber\\
{\cal J}_{\bf pp'kq'}^{\nu\mu s\gamma}&\overset{\rm def}{=}n_{\nu-}^{\uparrow}({\bf p})n_{\mu+}^{\uparrow}({\bf p}')n_{s-}^{\downarrow}({\bf k})
n_{\gamma+}^{\downarrow}({\bf q}')-n_{\nu+}^{\uparrow}({\bf p})n_{\mu-}^{\uparrow}({\bf p}')n_{s+}^{\downarrow}({\bf k})
n_{\gamma-}^{\downarrow}({\bf q}').
\end{align}
The high-frequency fluctuations come from the scattering processes from the upper to the lower band or the other way round via the background.
Note that the expressions in Eqs.~(\ref{lambda1}) and (\ref{lambda2}) resemble the structure of transition probability in time-dependent perturbation theory (see, e.g. Ref.~\cite{Sakurai_S}),
and the sinusoidal time dependence $\sin^2(\Delta_{\bf pp'qq'}^{s_1s_2s_3s_4} t/2)/(\Delta_{\bf pp'qq'}^{s_1s_2s_3s_4})^2$
determines the contribution of each scattering process in the time evolution.
One can find that the dependence $\sin^2(\omega t/2)/\omega^2$, as a function of $\omega$ for fixed $t$, has a major peak with the height $\propto t^2$ and the width
$\propto 1/t$~\cite{Sakurai_S}. Hence, after a summation, the parameters $\lambda^{\sigma}_{1,2}(t)$ are approximately linear in $t$, i.e., $\lambda^{\sigma}_{1,2}(t)\propto t$ (see Fig.~\ref{figS1}).

\begin{figure}
\includegraphics[width=0.58\textwidth]{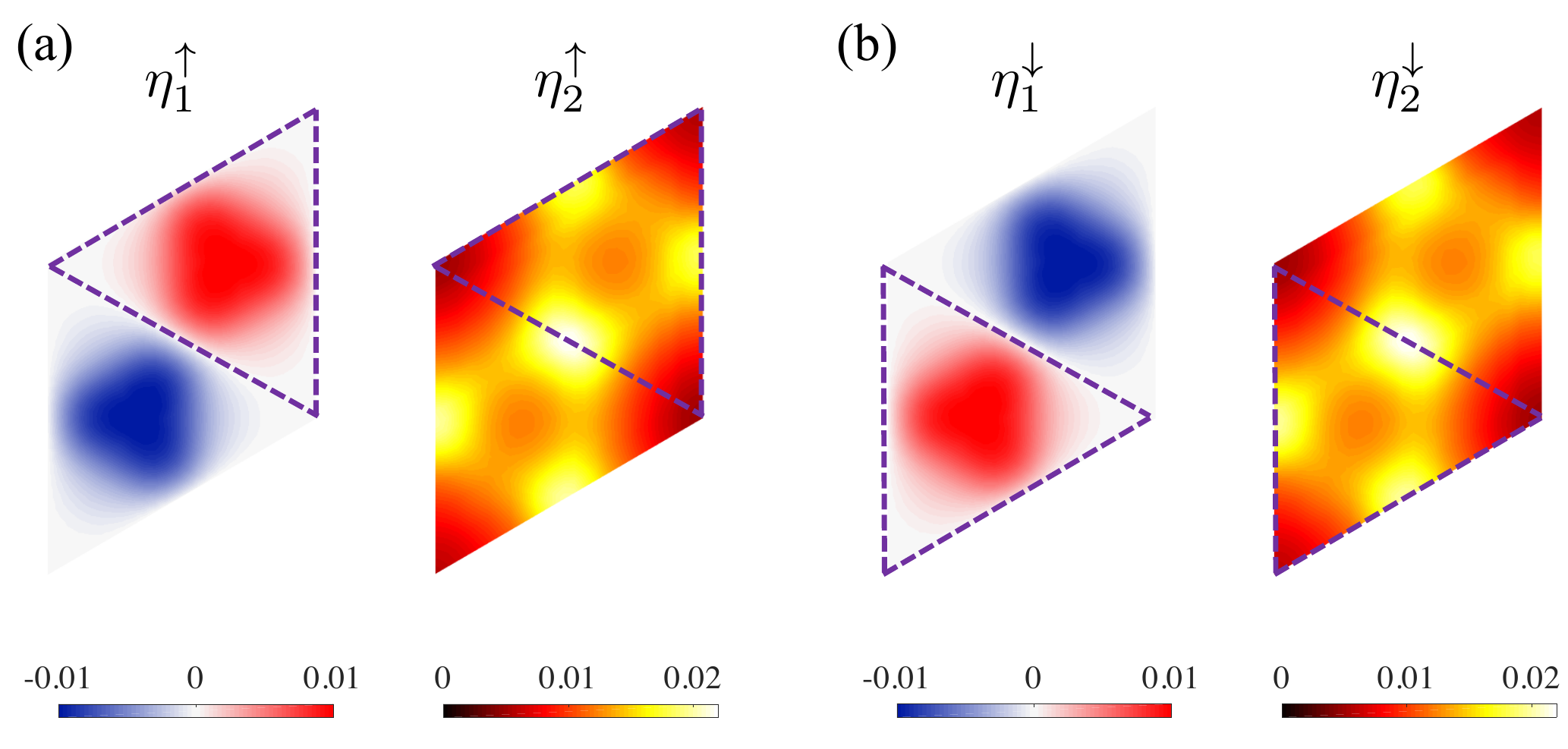}
\caption{The distribution of the damping factors $\eta_{1}^\sigma$ and heating
factors $\eta_{2}^\sigma$ for each spin with $t_2=0.3t_1$, $M=0$, $m_{\rm AF}\to\infty$ and the post-quench interaction $U=0.3t_1$. The dashed purple lines denote the noninteracting BIS.}\label{figS2}
\end{figure}

We define the pseudospin vector
\begin{equation}
{\bf S}_{\sigma}({\bf k},t)\overset{\rm def}{=}\frac{1}{2}\left(X_{\sigma}({\bf k},t),Y_{\sigma}({\bf k},t),Z_{\sigma}({\bf k},t)\right),
\end{equation}
and assume the equation of motion takes the form
\begin{equation}\label{EoMS}
\frac{d{\bf S}_{\sigma}(t)}{dt}={\bf S}_{\sigma}(t)\times2{\bf h}-\eta_1^\sigma{\bf S}_{\sigma}(t)\times\frac{d{\bf S}_{\sigma}(t)}{dt}-\eta_2^\sigma\frac{{\bf S}_{\sigma}(t)}{T_{\rm g}},
\end{equation}
where the first term in the right-hand side corresponds to the precessional motion, $\eta_{1}^\sigma$-term represents the damping effect, and
$\eta_{2}^\sigma$-term describes the heating.
According to Eqs.~(\ref{XYZ}-\ref{lambda2}), we obtain
\begin{align}\label{eta12}
\eta^\sigma_1({\bf k})=&2T_{\rm g}\left\{\frac{d\lambda^\sigma_{1}}{dt}\left[n_{+-}^\sigma({\bf k})-n_{--}^\sigma({\bf k})\right]-2\frac{d\lambda^\sigma_{2}}{dt}\right\},\nonumber\\
\eta^\sigma_2({\bf k})=&2T_{\rm g}\left\{2\frac{d\lambda^\sigma_{1}}{dt}n_{+-}^\sigma({\bf k})n_{--}^\sigma({\bf k})+\frac{d\lambda^\sigma_{2}}{dt}\left[n_{+-}^\sigma({\bf k})-n_{--}^\sigma({\bf k})\right]\right\}.
\end{align}
Note that $\eta^{\sigma}_{1,2}({\bf k})$ are approximated as time-independent in short time due to the linear time dependence of $\lambda^{\sigma}_{1,2}$ (Fig.~\ref{figS1}).
In Fig.~\ref{figS2}, we show the calculated results of $\eta_{1,2}^\sigma$ for $M=0$, $m_{\rm AF}\to\infty$, which
have a symmetrical (for $\eta_{2}^\sigma$) or antisymmetrical (for $\eta_{1}^\sigma$) distribution.

Finally, we discuss the reliability of our method.
It should be pointed out that secular terms may arise from our zeroth order approximation for the time evolution, e.g. in Eq.~(\ref{evolution}),
we take $H(l\to\infty)\approx H_0$. When $\Delta_{\bf p'pq'q}^{s_1s_2s_3s_4}=0$, the canonical generator (\ref{generator}) vanishes. Hence
the energy-diagonal contributions of $H_I$ cannot be erased by flow equations.  The perturbation solutions would fail on long-time scales.
This failure can be also indicated by the sinusoidal time dependence in Eqs.~(\ref{lambda1}) and (\ref{lambda2}). For a large $t$, the function
$\sin^2(\omega t/2)/\omega^2$ has a very narrow peak, and approximate energy conservation is required, which means the energy-diagonal contributions can not be neglected for a long time evolution.
Fortunately, we only need to focus on short-time pseudospin dynamics, from which the topology as well as magnetic order can be measured.
Furthermore, from the early stage dynamics, we can qualitatively analyze which interaction effect dominates
even for a relatively long time.

\section{IV. Detecting the topology}

In Ref.~\cite{Zhanglin2018_S}, we have developed a dynamical classification theory, which is applicable to noninteracting topological
systems and to the situation that the quench starts from a deep trivial regime. Here we first generalize the theory to the shallow
quench case, which corresponds to initializing finite magnetization in the interaction quench, and then discuss its feasibility in the current interacting system.

\subsubsection{A. Projection approach}

In this subsection, we consider the noninteracting system and generalize the dynamical classification theory in Ref.~\cite{Zhanglin2018_S}
to the situation that the pre-quench state is not completely polarized.
For a post-quench Hamitonian ${\cal H}({\bf k})={\bf h}({\bf k})\cdot{\bm \tau}$,
the spin texture reads ($i=x,y,z$)
\begin{align}
\overline{\langle\tau_i({\bf k})\rangle}=\lim_{T\to\infty}\frac{1}{T}\int_{0}^{t}d t{\rm Tr}[\rho_0e^{\ui {\cal H}t}\tau_{i}e^{-\ui {\cal H}t}]=\frac{h_{i}}{E_0^2}{\rm Tr}[\rho_0{\cal H}],
\end{align}
where $\rho_0$ is the density matrix of the initial state.
Although it is defined for an infinite time period,
the time average can be taken over several oscillations for each ${\bf k}$, and the results are unchanged.
The band inversion surfaces (BISs)  are defined as
\begin{align}
{\rm BIS}=\{\mathbf{k}\vert\overline{\langle\tau_{i}({\bf k})\rangle}=0\text{ for }i=x,y,z\}
\end{align}
This implies  that on the (noninteracting) BIS, the spin vector ${\bf S}({\bf k})\equiv\frac{1}{2}\left(\langle\tau_x\rangle,\langle\tau_y\rangle,\langle\tau_z\rangle\right)$
is perpendicular to the field ${\bf h}({\bf k})$, i.e., ${\bf S}({\bf k})\cdot{\bf h}({\bf k})=0$.
We denote
by $k_{\perp}$ the direction perpendicular to the contour ${\rm Tr}[\rho_0{\cal H}]$.
For the contours infinitely close to the BIS,
we have ${\rm Tr}[\rho_0{\cal H}]\simeq\pm k_{\perp}$ and the variation of $h_i$
is of order $\mathcal{O}(k_{\perp})$. Therefore, the directional
derivative on the BIS reads
\begin{align}\label{derivative_S}
\partial_{k_{\perp}}\overline{\langle\tau_i\rangle}=\lim_{k_{\perp}\to0}\frac{1}{2k_{\perp}}\left[\frac{h_i+\mathcal{O}(k_{\perp})}
{E_0^{2}+\mathcal{O}(k_{\perp})}k_{\perp}-\frac{h_i+\mathcal{O}(k_{\perp})}{E_0^{2}+\mathcal{O}(k_{\perp})}(-k_{\perp})\right]=\frac{h_i}{E_0^{2}}.
\end{align}

Without loss of generality, we consider quenching ${h}_z$.
When the initial state $\rho_0$ is fully polarized ($|h_z|\to\infty$ for $t<0$), the BIS conincides with the surfaces with $h_z({\bf k})=0$,
and $\partial_{k_{\perp}}\overline{\langle\tau_z\rangle}$ vanishes. Hence $\partial_{k_{\perp}}\overline{\langle{\bm\tau}\rangle}$ is a vector in the $x$-$y$ plane,
and the bulk topology is well defined by the winding of the spin-orbit (SO) field ${\bf h}_{\rm so}\equiv(h_y,h_x)$ along BISs, which is characterized by the dynamical field
$\partial_{k_{\perp}}\overline{\langle{\bm\tau}\rangle}=(\partial_{k_{\perp}}\overline{\langle\tau_y\rangle},\partial_{k_{\perp}}\overline{\langle\tau_x\rangle})$~\cite{Zhanglin2018_S}.
From the viewpoint of topological charges, which are located at ${\bf h}_{\rm so}({\bf k})=0$,
the winding of $\partial_{k_{\perp}}\overline{\langle{\bm\tau}\rangle}$
counts the total charges enclosed by BISs~\cite{Zhanglong2018_S}.
The BIS where ${\bf S}({\bf k})\cdot{\bf h}({\bf k})=0$ divides the charges into two categories: ${\bf S}({\bf k})\cdot{\bf h}({\bf k})>0$ and ${\bf S}({\bf k})\cdot{\bf h}({\bf k})<0$.
The winding of $\partial_{k_{\perp}}\overline{\langle{\bm\tau}\rangle}$ in fact characterizes the charges of the same category.

Now we consider the case that the initial state is not completely polarized, i.e., at $t=0$,
$\langle\tau_z({\bf k})\rangle=1$ (or $-1$) does not hold for all ${\bf k}$ but
$\langle\tau_z({\bf k})\rangle>0$ (or $<0$) does.
In this case, the topological charges enclosed
by BISs are unchanged. The reason is as follows:
First, by definition their locations are irrelevant to the initial state.
Second, the category, i.e., the condition ${\bf S}({\bf k})\cdot{\bf h}({\bf k})=\frac{1}{2}h_z({\bf k})\langle\tau_z({\bf k})\rangle>0$ or $<0$,
remains the same.
The BIS encloses the same charges as in the completely polarized case.
Note that the topological charges are characterized by the winding of ${\bf h}_{\rm so}$.
Although the vector $\partial_{k_{\perp}}\overline{\langle{\bm\tau}\rangle}$ is not in the $x$-$y$ plane in general, we can define the topological invariant by the winding of
a projected dynamical field $(\partial_{k_{\perp}}\overline{\langle\tau_y\rangle},\partial_{k_{\perp}}\overline{\langle\tau_x\rangle})$.
The dynamical field defined in the completely polarized case can be also regarded as a projection of $\partial_{k_{\perp}}\overline{\langle{\bm\tau}\rangle}$ but with
$\partial_{k_{\perp}}\overline{\langle\tau_z\rangle}=0$.

\subsubsection{B. Dynamical characterization in interacting systems}

In this subsection, we will show that the dynamical characterization theory discussed above is also applicable to the interacting Haldane model.
According to the results shown in Eqs.~(\ref{incoherent}-\ref{low_fuctuation}), the time-averaged pseudospin textures in the presence of interaction are
($i=x,y,z$)
\begin{align}
\overline{\langle\tau_i^\sigma\rangle}=\langle\tau_i^{\sigma(0)}\rangle+\overline{\langle\tau_i^{\sigma(l)}\rangle}=\frac{h_i}{E_0}\left[n_{+-}^\sigma({\bf k})-n_{--}^\sigma({\bf k})-\frac{d\lambda^\sigma_{2}({\bf k})}{dt}T\right],
\end{align}
where $T$ is the period over which the time average is taken and $\lambda^\sigma_{2}\propto U^2$ defined in Eq.~(\ref{lambda2}) represents the interaction shift.
Thus, the (interacting) BIS is determined by $\overline{\langle\tau^\sigma_i({\bf k})\rangle}=0$, which leads to
\begin{align}\label{interacting_BIS}
\delta n^\sigma_I({\bf k})\equiv n_{+-}^\sigma({\bf k})-n_{--}^\sigma({\bf k})-\frac{d\lambda^\sigma_{2}({\bf k})}{dt}T=0.
\end{align}
Note that in the interacting system, $k_\perp$ is defined to be perpendicular to the contour of $\delta n^\sigma_I({\bf k})$.
For the contours infinitely close to $\delta n^\sigma_I({\bf k}_0)=0$,  we have $\delta n^\sigma_I({\bf k}_0\pm\hat{e}_\perp k_{\perp})
\simeq\pm c_Ik_{\perp}/E_0$, with $c_I$ being a coefficient dependent on
$m_\sigma$, $U$ and $T$.
Therefore, we have
\begin{align}
\partial_{k_{\perp}}\overline{\langle\tau^\sigma_i\rangle}=\lim_{k_{\perp}\to0}\frac{1}{2k_{\perp}}\left[\frac{h_i}{E_0^{2}}\delta n^\sigma_I
({\bf k}_0+\hat{e}_\perp k_{\perp})-\frac{h_i}{E_0}\delta n^\sigma_I({\bf k}_0-\hat{e}_\perp k_{\perp})\right]=c_I\frac{h_i}{E_0^{2}},
\end{align}
which means the emergent gradient field $\partial_{k_{\perp}}\overline{\langle{\bm\tau}^\sigma\rangle}$ on the (interacting) BIS still
characterizes the vector field ${\bf h}({\bf k})$ despite of the interaction effect.
Due to the AF order, quenches for the two spins $\sigma=\uparrow\downarrow$ are along opposite directions.
Thus, according to Ref.~\cite{Zhanglin2018_S},
we define the projected dynamical fields on the BIS
${\bf g}_{\parallel}^\sigma({\bf k})=(g^\sigma_y,g^\sigma_x)$ with components given by
\begin{align}\label{dyfield_S}
g^\sigma_{y,x}({\bf k})=\pm\frac{1}{{\cal N}_k}\partial_{k_\perp}\overline{\langle\tau^\sigma_{y,x}\rangle}.
\end{align}
Here the sign $+$ (or $-$) is for $\sigma=\uparrow$ (or $\downarrow$) and ${\cal N}_k$ is the normalization factor.
The topological invariant is then defined by the winding of the projected dynamical field with $\sigma=\uparrow$ or $\downarrow$:
\begin{align}
w=\sum_j\frac{1}{2\pi}\int_{{\rm BIS}_j} \left[g_y^\sigma({\bf k})dg_x^\sigma({\bf k})-g_x^\sigma({\bf k})dg_y^\sigma({\bf k})\right].
\end{align}

\begin{figure}
\includegraphics[width=0.6\textwidth]{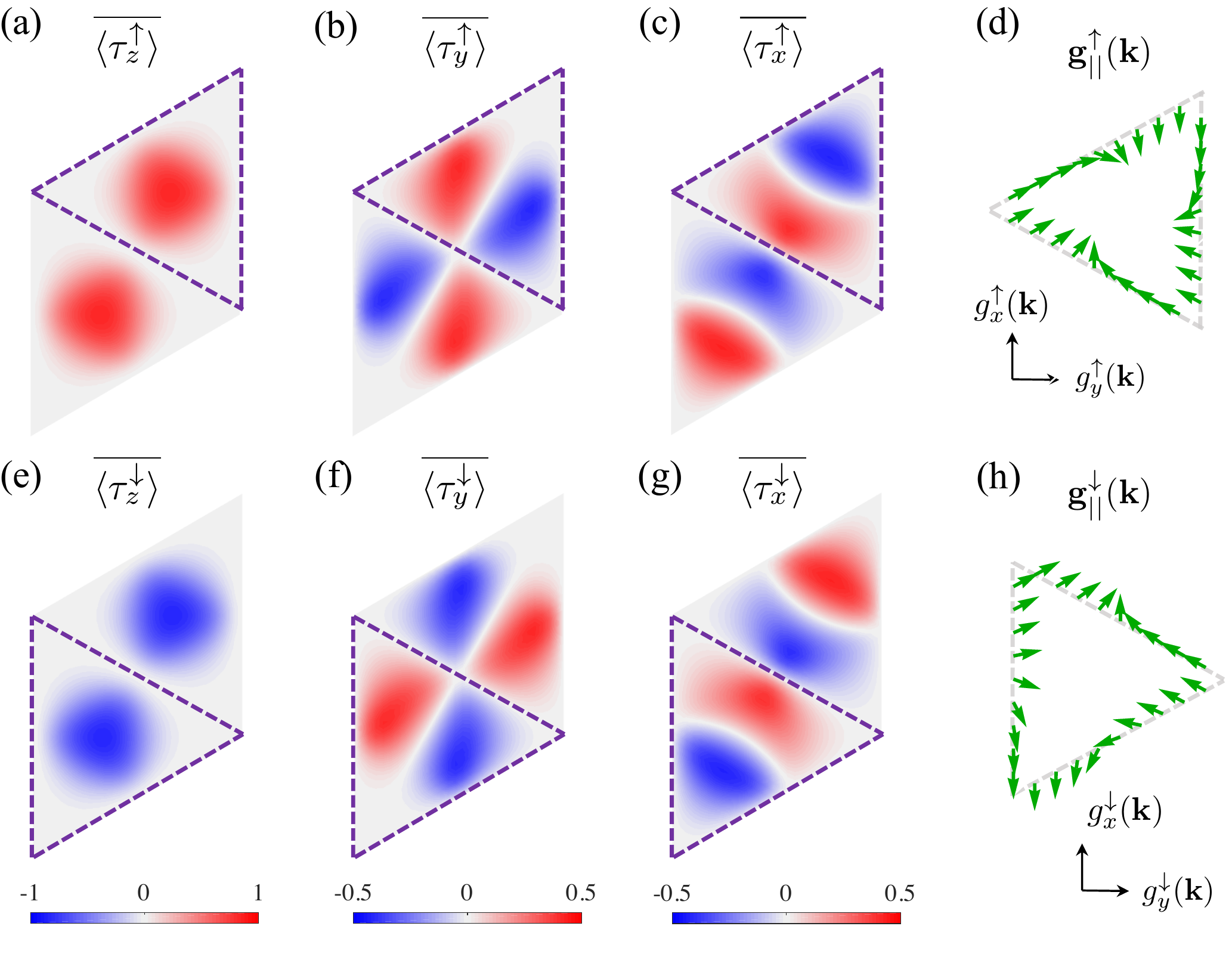}
\caption{
Dynamical classification of topology with completely polarization. Time-averaged pseudospin polarizations
$\overline{\langle\tau_{i}^\sigma({\bf k})\rangle}$ ($i=x,y,z$)
and the dynamical spin-texture fields ${\bf g}_{\parallel}^\sigma({\bf k})$ are shown. The dashed lines denotes the BISs.
The projected dynamical field on the BIS for either spin characterizes the topology with $C=1$ (d,h).
Here we take $t_2=0.3t_1$, $M=0$, $m_{\rm AF}\to\infty$ and the post-quench interaction $U=0.3t_1$.
The time average is taken over 10 times of oscillation period for each ${\bf k}$.
}\label{figS3}
\end{figure}

Here a special case is shown in Fig.~\ref{figS3} with $M=0$ and $m_{\rm AF}\to\infty$.
In this case, one can see that the damping factor $\eta^\sigma_1=-4T_{\rm g}d\lambda^\sigma_{2}/dt$ vanishes right on the
noninteracting BIS where $n_{--}({\bf k})=n_{+-}({\bf k})$ (Fig.~\ref{figS2}),
which is due to the exact cancelling of the two contributions in Eq~(\ref{lambda2}).
Thus, the BIS does not move in the presence of interaction.
Moreover, the distributions of $\eta^\sigma_1$ are antisymmetrical.
As shown in Fig.~\ref{figS3}, the time-averaged textures
$\overline{\langle\tau_{i}^\sigma({\bf k})\rangle}$ take the same distributions as in the noninteracing case, except
for a small reduction of polarization values. The time averages are taken over 10 times of oscillation period for each ${\bf k}$.
An example of a general case with finite $m_{\rm AF}$ is discussed in Fig.~3 of the main text.

\section{V. Measuring the magnetic order}
We aim to obtain the magnetic order $m_{\sigma}$ by measuring the pseudospin dynamics.
In the presence of interaction, the BIS is given by Eq.~(\ref{interacting_BIS}). Here we assume $|d\lambda^\sigma_{2}/dt|T\ll1$.
Since
\begin{align}
n_{+-}^\sigma({\bf k})=\frac{1}{2}-\frac{E_0^2+m_\sigma h_z}{2E_0E_0^\sigma},\quad
n_{--}^\sigma({\bf k})=\frac{1}{2}+\frac{E_0^2+m_\sigma h_z}{2E_0E_0^\sigma},
\end{align}
where $E_0^\sigma\equiv\sqrt{E_0^2+2m_\sigma h_z+m_\sigma^2}$,
the BIS can be alternatively interpreted as the momenta
satisfying
\begin{equation}\label{BIS_i}
E_0^2({\bf k})+m_\sigma h_z({\bf k})=-\frac{d\lambda^\sigma_{2}({\bf k})}{dt}TE_0({\bf k})E_0^\sigma({\bf k}).
\end{equation}
Note that when $h_x=h_y=0$, the above equation becomes $(1+Td\lambda^\sigma_{2}/dt)(h_z^2+m_\sigma h_z)=0$, which fails to hold for $|m_\sigma|>|h_z|$ except that $h_z=0$.
That is to say when we consider the quench from a trivial phase ($|m_\sigma|>|h_z|$), the BIS would not move across a charge where ${\bf h}_{\rm so}=0$ except that topological phase transition occurs.
Furthermore, half of the amplitude in the early time
 $Z_{0}^\sigma({\bf k})\overset{\rm def}{=}\langle\tau_z^\sigma({\bf k},t=0)\rangle$
reads
\begin{equation}\label{Z0_S}
Z^\sigma_0({\bf k})=Z_\sigma^{(0)}({\bf k})+Z_\sigma^{(c)}({\bf k},0)=\frac{h_z({\bf k})}{E_0({\bf k})}\frac{d\lambda^\sigma_{2}({\bf k})}{dt}T-\frac{E_0^2({\bf k})-h_z^2({\bf k})}{E_0^2({\bf k})}\frac{m_\sigma}{E_0^\sigma({\bf k})}
\end{equation}
on the BIS.
Equations~(\ref{BIS_i}) and (\ref{Z0_S}) provide two relations for the derivation of the magnetic order $m_\sigma$. We regard
the interaction effect as a perturbation and approximate $m_\sigma$ and $h_z$ to the first order of $\varepsilon\equiv T d\lambda^\sigma_{2}/dt$, i.e.
$m_\sigma=m_\sigma^{(0)}+\varepsilon m_\sigma^{(1)}$ and $h_z=h_z^{(0)}+\varepsilon h_z^{(1)}$.
We then have
\begin{subequations}
\begin{align}
E_0^2+m_\sigma^{(0)} h_z^{(0)}&=0, \label{Eq_Sa} \\
m_\sigma^{(1)} h_z^{(0)}+m_\sigma^{(0)} h_z^{(1)}&=-E_0\sqrt{m_\sigma^{(0)2}-E_0^2}, \label{Eq_Sb} \\
Z_{0}^{\sigma}\sqrt{m_\sigma^{(0)2}-E_0^2}&=-\left(1-\frac{h_z^{(0)2}}{E_0^2}\right)m_\sigma^{(0)}, \label{Eq_Sc} \\
\frac{Z_{0}^{\sigma}}{\sqrt{m_\sigma^{(0)2}-E_0^2}}\left(m_\sigma^{(1)} h_z^{(0)}+m_\sigma^{(0)} h_z^{(1)}+m_\sigma^{(1)}
m_\sigma^{(0)}\right)&=\frac{h_z^{(0)}}{E_0}\sqrt{m_\sigma^{(0)2}-E_0^2}+
m_\sigma^{(0)}\frac{2h_z^{(1)}h_z^{(0)}}{E_0^2}-m_\sigma^{(1)}\left(1-\frac{h_z^{(0)2}}{E_0^2}\right). \label{Eq_Sd}
\end{align}
\end{subequations}
From Eqs.~(\ref{Eq_Sa}) and (\ref{Eq_Sc}), we obtain
\begin{align}
m_\sigma^{(0)}=-{\rm sgn}(Z^\sigma_0)\frac{E_0}{\sqrt{1-Z^{\sigma\,2}_0}}, \quad\quad
h_z^{(0)}={\rm sgn}(Z^\sigma_0)E_0\sqrt{1-Z^{\sigma\,2}_0}.
\end{align}
Substituting the results into Eqs.~(\ref{Eq_Sb}) and (\ref{Eq_Sd}) leads to
\begin{align}
m_\sigma^{(1)}=0, \quad\quad h_z^{(1)}=E_0Z_0^\sigma.
\end{align}
Finally, to the second order of $U$, we have
\begin{equation}
m_\sigma T_0=-{\rm sgn}(Z^\sigma_0)\frac{\pi}{\sqrt{1-Z^{\sigma\,2}_0}},  \  \mbox{with}\ T_0=\pi/E_0(\bold k),
\end{equation}
which is the universal scaling behavior immune to the interaction.
The AF and charge orders are finally given by $m_{\rm AF}=(m_\downarrow-m_\uparrow)/2$ and
$m_{\rm C}=(m_\uparrow+m_\downarrow)/2$.

\section{VI. Results beyond mean field theory}
In this section, we examine the beyond-mean-field (BMF) effect of the initial state. To this end, we
take a Gutzwiller wave function~\cite{Gutzwiller_S}, instead of a mean-field ground state,  to describe the AF order phase, and examine
the corrections to our results.

\subsubsection{A. Gutzwiller wave function}
We take the Gutzwiller ansatz (see, e.g., Refs~\cite{Li1993_S,Eichenberger2007_S} and references therein)
\begin{align}
|\Psi_{\rm G}\rangle=\prod_i(1-\alpha A_i)(1-\alpha B_i)|\Psi_{\rm MF}\rangle,
\end{align}
where $\alpha $ is the variational parameter with $0\leq \alpha \leq 1$. Here we have denoted $A_i\equiv a^\dagger_{i\uparrow}a^\dagger_{i\downarrow}a_{i\downarrow}a_{i\uparrow} $ and $B_i\equiv b^\dagger_{i\uparrow}b^\dagger_{i\downarrow}b_{i\downarrow}b_{i\uparrow}$, and $|\Psi_{\rm MF}\rangle$ is the mean-field ground state. The $\alpha $-terms suppress the double occupation of the mean-field ground state.
To explore the correlation effect, we keep the leading-order terms and obtain that
\begin{align}
|\Psi_{\rm G}\rangle&\approx\left[1-\alpha\sum_i(A_i+B_i)\right]|\Psi_{\rm MF}\rangle.
\end{align}
After transforming into the momentum space, we have the {\it unnormalized} wave function
\begin{align}\label{GWFk}
|\Psi_{\rm G}\rangle\approx\left(1-\alpha\sum_{\substack{{\bf p'pq'q}\\ s_1s_2s_3s_4}}\delta^{\bf p+q}_{\bf p'+q'}\Lambda_{\bf p'pq'q}^{s_1s_2s_3s_4}c_{{\bf p'},s_1\uparrow}^\dagger c_{{\bf p},s_2\uparrow}c_{{\bf q'},s_3\downarrow}^\dagger c_{{\bf q},s_4\downarrow}\right)|\Psi_{\rm MF}\rangle,
\end{align}
with $\Lambda_{\bf p'pq'q}^{s_1s_2s_3s_4}$ given by Eq.~(\ref{Lambda}).
One can see that the constructed wave function is in fact a superposition of the mean-field state and a series of excited states via single scattering. Thus the correlation effects are explicitly incorporated in the Gutzwiller mang-body ground state.
The initial state $|\Psi_{\rm MF}\rangle$
is updated by diagonalizing the mean-field Hamiltonian with the renormalized order parameters
$m_\uparrow=m_{\rm C}-m_{\rm AF}$ and $m_\downarrow=m_{\rm C}+m_{\rm AF}$, where
\begin{align}
m_{\rm AF}\equiv&\langle b^\dagger_{i\uparrow}b_{i\uparrow}-b^\dagger_{i\downarrow}b_{i\downarrow}\rangle U_{\rm in}/2=-\langle a^\dagger_{i\uparrow}a_{i\uparrow}-a^\dagger_{i\downarrow}a_{i\downarrow} \rangle U_{\rm in}/2,\nonumber\\
m_{\rm C}\equiv&\langle a^\dagger_{i\uparrow}a_{i\uparrow}+a^\dagger_{i\downarrow}a_{i\downarrow}- b^\dagger_{i\uparrow}b_{i\uparrow}- b^\dagger_{i\downarrow}b_{i\downarrow}\rangle U_{\rm in}/4.
\end{align}
Here $\langle\cdot\rangle=\langle\Psi_{\rm G}|\cdot|\Psi_{\rm G}\rangle$ is now calculated based on $|\Psi_{\rm G}\rangle$. We note that to solve the Gutzwiller ground state of the initial Hamiltonian, one needs to determine the variational parameter $\alpha$ and
the wave function $|\Psi_{\rm G}\rangle$ with renormalized $m_{\uparrow,\downarrow}$ iteratively until the energy $\langle H\rangle_{\rm G}$ is minimized. For the purpose of the present study, we only need to formally write down the ground state in the Gutzwiller form, and show that the quench dynamics can extract the information of the ground state.

\subsubsection{B. Correlations and flow equations}

Compared with Eq.~(\ref{correlation_MF}), we have nonzero correlations with respect to the Gutzwiller wave function ($\mu,\nu=\pm$, ${\bf k}_1\neq{\bf k}_2$)
\begin{align}\label{crltn_G}
\langle c_{{\bf k},\mu\uparrow}^\dagger c_{{\bf k},\nu\uparrow}\rangle_{\rm G}\simeq&f_{\mu-}^{\uparrow\,*}({\bf k})f_{\nu-}^{\uparrow}({\bf k})-\alpha {\cal C}_{\mu\nu}^{\uparrow}({\bf k})+{\cal O}(\alpha^2),
\nonumber\\
\langle c_{{\bf k},\mu\downarrow}^\dagger c_{{\bf k},\nu\downarrow}\rangle_{\rm G}\simeq& f_{\mu-}^{\downarrow\,*}({\bf k})f_{\nu-}^{\downarrow}({\bf k})-\alpha {\cal C}_{\mu\nu}^{\downarrow}({\bf k})+{\cal O}(\alpha^2),
\nonumber\\
\langle c_{{\bf k},\mu\sigma}c_{{\bf k},\nu\sigma}^\dagger \rangle_{\rm G}=&\delta_{\mu}^{\nu}-\langle c_{{\bf k},\nu\sigma}^\dagger c_{{\bf k},\mu\sigma}\rangle,
\end{align}
where the BMF corrections
\begin{align}
{\cal C}_{\mu\nu}^{\sigma}({\bf k})=&\sum_{{\bf q},s_2s_3s_4}\left[\Lambda_{\bf kkqq}^{\nu s_2s_3s_4}f_{\mu-}^{\sigma\,*}({\bf k})f_{s_2-}^{\sigma}({\bf k})f_{s_3-}^{\bar\sigma\,*}({\bf q})f_{s_4-}^{\bar\sigma}({\bf q})+\Lambda_{\bf kkqq}^{s_2\mu s_4s_3}f_{s_2-}^{\sigma\,*}({\bf k})f_{\nu-}^{\sigma}({\bf k})f_{s_4-}^{\bar\sigma\,*}({\bf q})f_{s_3-}^{\bar\sigma}({\bf q})\right] \nonumber\\
&+2f_{\mu-}^{\sigma\,*}({\bf k})f_{\nu-}^{\sigma}({\bf k})\sum_{\substack{{\bf q,p\neq k}\\ s_1s_2s_3s_4}}\Re\left[\Lambda_{\bf ppqq}^{s_1s_2s_3s_4}f_{s_1-}^{\sigma\,*}({\bf p})f_{s_2-}^{\sigma}({\bf p})f_{s_3-}^{\bar\sigma\,*}({\bf q})f_{s_4-}^{\bar\sigma}({\bf q})\right].
\end{align}
Here we denote $\bar\uparrow=\downarrow$ and $\bar\downarrow=\uparrow$.
Based on the above results, one can easily check that
the generator $\eta(l)$ [see Eq.~(\ref{generator})] and the ansatzes for creation operators [Eq.~(\ref{Al_Flow})] remain unchanged.
Moreover, the leading-order flow equations take the same forms as Eqs.~(\ref{solu_hM}) and (\ref{solu_gW}), except that the parameters
$F_{\bf p'qq'}^{s_2s_3s_4,\mu\gamma\nu}$ and $G_{\bf pp'q'}^{s_1s_2s_4,\nu\mu\gamma}$ should be, respectively, replaced by $\widetilde{F}_{\bf p'qq'}^{s_2s_3s_4,\mu\gamma\nu}$
and $\widetilde{G}_{\bf pp'q'}^{s_1s_2s_4,\nu\mu\gamma}$, where
\begin{subequations}
\begin{align}
\widetilde{F}_{\bf p'qq'}^{s_2s_3s_4,\mu\gamma\nu}=&\left[f_{s_2+}^{\uparrow}({\bf p}')f_{\mu+}^{\uparrow\,*}({\bf p}')+g{\cal C}_{\mu s_2}^{\uparrow}({\bf p'})\right]
\left[f_{s_3-}^{\downarrow\,*}({\bf q})f_{\gamma-}^{\downarrow}({\bf q})-\alpha {\cal C}_{s_3\gamma}^{\downarrow}({\bf q})\right]
\left[f_{s_4+}^{\downarrow}({\bf q}')f_{\nu+}^{\downarrow\,*}({\bf q}')+\alpha {\cal C}_{\nu s_4}^{\downarrow}({\bf q'})\right]\nonumber\\
&+\left[f_{s_2-}^{\uparrow}({\bf p}')f_{\mu-}^{\uparrow\,*}({\bf p}')-\alpha {\cal C}_{\mu s_2}^{\uparrow}({\bf p'})\right]
\left[f_{s_3+}^{\downarrow\,*}({\bf q})f_{\gamma+}^{\downarrow}({\bf q})+\alpha {\cal C}_{s_3\gamma}^{\downarrow}({\bf q})\right]
\left[f_{s_4-}^{\downarrow}({\bf q}')f_{\nu-}^{\downarrow\,*}({\bf q}')-\alpha {\cal C}_{\nu s_4}^{\downarrow}({\bf q}')\right],\\
\widetilde{G}_{\bf pp'q'}^{s_1s_2s_4,\nu\mu\gamma}=&\left[f_{s_1-}^{\uparrow\,*}({\bf p})f_{\nu-}^{\uparrow}({\bf p})-g{\cal C}_{s_1\nu}^{\uparrow}({\bf p})\right]
\left[f_{s_2+}^{\uparrow}({\bf p}')f_{\mu+}^{\uparrow\,*}({\bf p}')+\alpha {\cal C}_{\mu s_2}^{\uparrow}({\bf p}')\right]
\left[f_{s_4+}^{\downarrow}({\bf q}')f_{\gamma+}^{\downarrow\,*}({\bf q}')+\alpha {\cal C}_{\gamma s_4}^{\downarrow}({\bf q}')\right]\nonumber\\
&+\left[f_{s_1+}^{\uparrow\,*}({\bf p})f_{\nu+}^{\uparrow}({\bf p})+\alpha {\cal C}_{s_1\nu}^{\uparrow}({\bf p})\right]
\left[f_{s_2-}^{\uparrow}({\bf p}')f_{\mu-}^{\uparrow\,*}({\bf p}')-\alpha {\cal C}_{\mu s_2}^{\uparrow}({\bf p}')\right]
\left[f_{s_4-}^{\downarrow}({\bf q}')f_{\gamma-}^{\downarrow\,*}({\bf q}')-\alpha {\cal C}_{\gamma s_4}^{\downarrow}({\bf q}')\right].
\end{align}
\end{subequations}
Similar to Eq.~(\ref{FG_simplified}),
we can further simplify the calculations by only taking into account the major contribution, i.e.
\begin{align}
\widetilde{F}_{\bf p'qq'}^{s_2s_3s_4,\mu\gamma\nu}&\simeq\delta_{s_2}^{\mu}\delta_{s_3}^{\gamma}\delta_{s_4}^{\nu}\widetilde{\cal F}_{\bf p'qq'}^{\mu\gamma\nu},\nonumber\\
\widetilde{G}_{\bf pp'q'}^{s_1s_2s_4,\nu\mu\gamma}&\simeq\delta_{s_1}^{\nu}\delta_{s_2}^{\mu}\delta_{s_4}^{\gamma}\widetilde{\cal G}_{\bf pp'q'}^{\nu\mu\gamma},
\end{align}
with
$\widetilde{\cal F}_{\bf p'qq'}^{\mu\gamma\nu}
\simeq {\cal F}_{\bf p'qq'}^{\mu\gamma\nu}+\alpha {\cal C}_{\mu\mu}^{\uparrow}({\bf p'})[n_{\gamma-}^{\downarrow}({\bf q})-n_{\nu-}^{\downarrow}({\bf q}')]
+\alpha {\cal C}_{\gamma\gamma}^{\downarrow}({\bf q})[n_{\mu-}^{\uparrow}({\bf p}')-n_{\nu+}^{\downarrow}({\bf q}')]
-\alpha {\cal C}_{\nu \nu}^{\downarrow}({\bf q'})[n_{\mu-}^{\uparrow}({\bf p}')-n_{\gamma-}^{\downarrow}({\bf q})]+{\cal O}(\alpha^2)$ and
$\widetilde{\cal G}_{\bf pp'q'}^{\nu\mu\gamma}\simeq {\cal G}_{\bf pp'q'}^{\nu\mu\gamma}+\alpha {\cal C}_{\nu\nu}^{\uparrow}({\bf p})[n_{\mu-}^{\uparrow}({\bf p}')-n_{\gamma+}^{\downarrow}({\bf q}')]
+\alpha {\cal C}_{\mu\mu}^{\uparrow}({\bf p}')[n_{\nu-}^{\uparrow}({\bf p})-n_{\gamma-}^{\downarrow}({\bf q}')]
+\alpha {\cal C}_{\gamma\gamma}^{\downarrow}({\bf q}')[n_{\nu-}^{\uparrow}({\bf p})-n_{\mu-}^{\uparrow}({\bf p}')]+{\cal O}(\alpha^2)$. With these results we conclude that the forms of flow equations remain the same,
while the BMF corrections only modify the parameters that depend on the initial momentum distribution.
Therefore, we find that the equation of motion keeps the same form as Eq.~\eqref{EoMS}, but the parameters are corrected by the interactions. Also, the initial spin length at each single particle momentum can be less than one, since in the initial Gutzwiller ground state the different momentum states are correlated.

\subsubsection{C. Universal scaling on the BIS}

From the results above, one conclude that the quench dynamics is governed by the equation of motion of the same form as Eq.~(\ref{EoMS}) but with modified parameters. With this we can expect that the emergent topology of the quench dynamics is not affected by the corrections, as long as the initial ground state is topologically trivial.
We then focus on the BMF correction to the universal scaling. The BIS is determined by
\begin{align}
\overline{\langle\tau_i^\sigma\rangle}=\langle\tau_i^{\sigma(0)}\rangle+\overline{\langle\tau_i^{\sigma(l)}\rangle}=\frac{h_i}{E_0}\left[n_{+-}^\sigma({\bf k})-n_{--}^\sigma({\bf k})-\alpha {\cal C}_{++}^{\sigma}({\bf k})+\alpha {\cal C}_{--}^{\sigma}({\bf k})-\frac{d\widetilde{\lambda}^\sigma_{2}({\bf k})}{dt}T\right]=0,
\end{align}
where $T$ denotes the evolution time, the mean-field density
\begin{align}
n_{\pm-}^\sigma({\bf k})=\frac{1}{2}\mp\frac{E_0^2+m_\sigma h_z}{2E_0\widetilde{E}_0^\sigma}
\end{align}
with $\widetilde{E}_0^\sigma\equiv\sqrt{E_0^2+2m_\sigma h_z+m_\sigma^2}$, and $\widetilde{\lambda}^\sigma_{2}({\bf k})$ are defined as
in Eqs.~(\ref{lambda2}) and (\ref{IJ_MF}) with $n^\sigma_{s\pm}({\bf k})$ being replaced by $n^\sigma_{s\pm}({\bf k})\pm \alpha {\cal C}_{ss}^{\sigma}({\bf k})$.
Since
\begin{align}
{\cal C}_{++}^{\sigma}({\bf k})-{\cal C}_{--}^{\sigma}({\bf k})&\simeq 2\zeta [n_{+-}^{\sigma}({\bf k})-n_{--}^{\sigma}({\bf k})],\nonumber\\
{\cal C}_{+-}^{\sigma}({\bf k})+{\cal C}_{-+}^{\sigma}({\bf k})&\simeq 2\zeta [f_{+-}^{\sigma\,*}({\bf k})f_{--}^{\sigma}({\bf k})+f_{--}^{\sigma\,*}({\bf k})f_{+-}^{\sigma}({\bf k})],
\end{align}
where we define
$\zeta\overset{\rm def}{=}\sum_{\substack{{\bf q,p}\\ s_1s_2s_3s_4}}\Re\left[\Lambda_{\bf ppqq}^{s_1s_2s_3s_4}f_{s_1-}^{\uparrow\,*}({\bf p})f_{s_2-}^{\uparrow}({\bf p})f_{s_3-}^{\downarrow\,*}({\bf q})f_{s_4-}^{\downarrow}({\bf q})\right]$,
we then know that on the BIS,
\begin{align}
E_0^2({\bf k})+m_\sigma h_z({\bf k})=-\frac{d\widetilde{\lambda}^\sigma_{2}({\bf k})}{dt}TE_0({\bf k})\widetilde{E}_0^\sigma({\bf k})/(1-2\alpha \zeta),
\end{align}
and half of the amplitude reads
\begin{equation}\label{Z0_S}
Z^\sigma_0({\bf k})=\frac{h_z({\bf k})}{E_0({\bf k})}\frac{d\widetilde{\lambda}^\sigma_{2}({\bf k})}{dt}T-\frac{E_0^2({\bf k})-h_z^2({\bf k})}{E_0^2({\bf k})}\frac{m_\sigma}{E_0^\sigma({\bf k})}(1-2\alpha \zeta).
\end{equation}
If we define $\widetilde{Z}^\sigma_0\equiv Z^\sigma_0/(1-2\alpha \zeta)$, we still have the universal scaling for quench dynamics on the BIS
\begin{equation}\label{scalingSM}
f(m_\sigma)=-\frac{{\rm sgn}(\widetilde{Z}^\sigma_0)}{g(\widetilde{Z}^\sigma_0)}+{\cal O}(U^4),
\end{equation}
where $f(m_\sigma)=m_\sigma T_0$ and $g(\widetilde{Z}^\sigma_0)=\sqrt{1-\widetilde{Z}_0^{\sigma\,2}}/\pi$. Note that the unknown constant $\alpha \zeta$ can be determined by measuring $Z^\sigma_0$ and $T_0$ of the dynamics at any two points on BIS, which satisfy the same scaling in the above Eq.~\eqref{scalingSM}.



\end{document}